\documentclass[aoas,preprint]{imsart}
\RequirePackage[authoryear]{natbib}

\usepackage{amsthm,amsmath}
\usepackage{amssymb}
\usepackage{graphicx}
\usepackage{enumitem}
\usepackage{comment}
\usepackage{array}
\usepackage{subcaption}
\usepackage{multirow}
\usepackage{tabularx}
\usepackage{arydshln}
\usepackage{qtree}

\usepackage{algorithm}

\RequirePackage[colorlinks,citecolor=blue,urlcolor=blue]{hyperref}

\setattribute{journal}{name}{} 

\startlocaldefs

\newcommand{\h}{\hat}

\newcommand\clearrow{\global\let\rowmac\relax}
\clearrow

\DeclareMathOperator{\logit}{logit}
\DeclareMathOperator{\cov}{cov}
\DeclareMathOperator{\var}{var}

\newcommand{\ex}{\mathbb{E}}

\newcommand{\pinv}{\dagger}
\newcommand{\T}{T}
\newcommand{\Normal}{\mathcal{N}}

\newcolumntype{L}[1]{>{\raggedright\let\newline\\\arraybackslash\hspace{0pt}}m{#1}}
\newcolumntype{C}[1]{>{\centering\let\newline\\\arraybackslash\hspace{0pt}}m{#1}}
\newcolumntype{R}[1]{>{\raggedleft\let\newline\\\arraybackslash\hspace{0pt}}m{#1}}

\usepackage{algpseudocode}
\algrenewcommand\textproc{}

\DeclareRobustCommand\citepos
{\begingroup
 \let\NAT@nmfmt\NAT@posfmt
 \NAT@swafalse\let\NAT@ctype\z@\NAT@partrue
 \@ifstar{\NAT@fulltrue\NAT@citetp}{\NAT@fullfalse\NAT@citetp}}

\let\NAT@orig@nmfmt\NAT@nmfmt
\def\NAT@posfmt#1{\NAT@orig@nmfmt{#1's}}

\endlocaldefs

\begin{document}

\begin{frontmatter}

\title{Fitting a deeply-nested hierarchical model \\
to a large book review dataset \\
using a moment-based estimator}
\runtitle{Fitting a deeply-nested hierarchical model}

\author{\fnms{Ningshan} \snm{Zhang}\thanksref{m1}\ead[label=e1]{nzhang@stern.nyu.edu}},
\author{\fnms{Kyle} \snm{Schmaus}\thanksref{m2}\ead[label=e2]{kschmaus@stichfix.com}}
\and
\author{\fnms{Patrick O.} \snm{Perry}\corref{}\thanksref{m1}\ead[label=e3]{pperry@stern.nyu.edu}}
\affiliation{New York University\thanksmark{m1} \\
and \\
Stitch Fix\thanksmark{m2}}

\address{
Information, Operations, \\
\phantom{Info} and Management Sciences Department \\
Stern School of Business \\
New York University \\
44 West 4th St. \\
New York City, NY 10012 \\
\printead{e1} \\
\phantom{E-mail:} \printead*{e3}
}

\address{
Stitch Fix \\
San Francisco, CA \\
\printead{e2}
}

\runauthor{Zhang et al.}

\begin{abstract}
We consider a particular instance of a common problem in recommender systems: using a database of book reviews to inform user-targeted recommendations. In our dataset, books are categorized into genres and sub-genres. To exploit this nested taxonomy, we use a hierarchical model that enables information pooling across across similar items at many levels within the genre hierarchy. The main challenge in deploying this model is computational: the data sizes are large, and fitting the model at scale using off-the-shelf maximum likelihood procedures is prohibitive. To get around this computational bottleneck, we extend a moment-based fitting procedure proposed for fitting single-level hierarchical models to the general case of arbitrarily deep hierarchies. This extension is an order of magnetite faster than standard maximum likelihood procedures. The fitting method can be deployed beyond recommender systems to general contexts with deeply-nested hierarchical generalized linear mixed models.
\end{abstract}



\end{frontmatter}

\section{Introduction}
\label{sec:intro}

\begin{figure}[ht!]
     \centering
     \includegraphics[width=\textwidth]{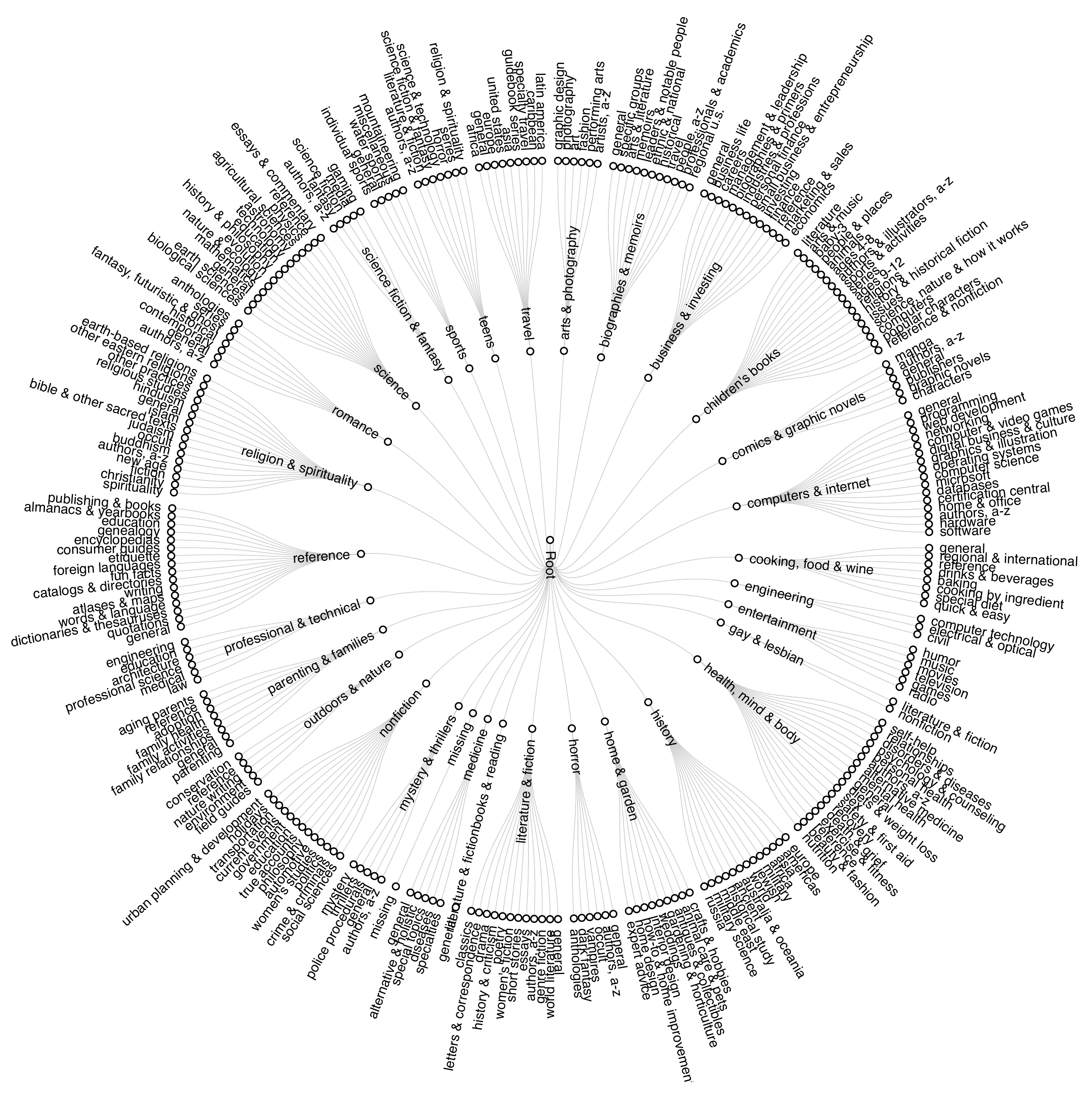}
     \caption{First two levels of the book genre hierarchy.}
\label{fig:2lpie}
\end{figure}

Given a dataset of  books, users, and user reviews of those books, consider the
problem of recommending books to users. This problem is a specific instance of
the recommender system problem common in commercial
applications~\citep{Adom05}. In our context the following data are available:
\begin{itemize}
    \item A collection of 38,659 books, each with an author title, genre,
        subgenre, and sub-subgenre, a taxonomy scraped
        from \texttt{amazon.com} by \citet{Mcau15}.
        Fig.~\ref{fig:2lpie} shows the first two levels of the book genre hierarchy.

    \item A set of 157,638 ratings of the books made by 38,085 users,
        taken from The Book Crossing Dataset, an anonymized collection of 
        book reviews harvested from \texttt{bookcrossing.com}
        by \citet{Zieg05}.

    \item User age and location (continent), included with the Book
        Crossing Dataset.
\end{itemize}

\noindent
Appendix~\ref{app:bookxdata} gives descriptive statistics for the dataset, including
descriptions of how the ratings are distributed between the user age groups
and continents. Most ratings are from users in the United States, aged
20--40 years.

Others have built recommender systems for the Book Crossing Dataset
\citep{4669763,Agarwal:2010:FMF:1718487.1718499,Zhang:2010:MCF:3023549.3023635}.
Our application is unique in that we will attempt to leverage the book genre
hierarchy to improve recommendations.

Rather than solving the book recommendation problem directly, we will attempt to
solve a proxy problem: for each book-user pair, predict whether the user would
like the book if he or she rated it.  We can use the solution to the proxy
problem to solve the original problem by recommending books with the highest
predicted ``like'' probabilities. This proxy approach to the recommendation
system is common~\citep{Ansari2000,Adom05}; it ignores information inherent in a user's
selection of which books to review, but despite this it often gives reasonable
downstream results.

One strategy for solving a recommender system problem is the content-based
approach, using user attributes together with book-specific parameters to make
recommendations. Another strategy is the collaborative approach,
recommending books that are liked by similar users.  We prefer instead a
variant of the hierarchical-model-based approach advocated by
\citet{Condliff1999} and \citet{Ansari2000} that combines the content-based
and collaborative strategies.

The simplest form of the hierarchical model approach is with a flat item
hierarchy, with each item sitting directly under the root.
In our context, the flat model would assign a random effect
vector~$u_i$ to each book~$i$ that relates the popularity of the book to user-
and context-specific covariate vector, along with a fixed effect vector
$\beta$ that relates book popularity to another covariate vector (possibly the
same). 
For a particular user review of a book~$i$, let $y$ be a binary indicator of
whether the user liked the book, and let $x_0$ and $x_1$ be the user-context
covariate vectors associate fixed and random effects, respectively. The link
between the effects, the covariates and the response is
\begin{equation}\label{eqn:flat-link}
    \logit \Pr(y = 1 \mid u_i)
    = \beta^\T x_0 + u_i^T x_1.
\end{equation}
The random effect vectors are independent multivariate normal random
vectors with mean $0$ and covariance matrix $\Sigma_1$:
\begin{equation}\label{eqn:flat-pop}
    u_i \sim \Normal(0, \Sigma_1).
\end{equation}
Eq.~\eqref{eqn:flat-link} demonstrates the content-based aspect of the model,
where user attributes ($x_0$ and $x_1$) are linked to preferences.
Eq.~\eqref{eqn:flat-pop} introduces the collaborative-based features:
books with abundant data will have strongly-identified random effects $u_i$;
others will have posterior means (conditional on the available review data)
determined in part by the effects of similar items, through the covariance
matrix~$\Sigma_1$.

In our application, we have a richer hierarchy, with books nested under
author, sub-subgenre, subgenere, and genre. We can exploit this hierarchy
in the model formulation by allowing for a random effect vector at each
node in the hierarchy, not just at the leaves. Doing so allows for
information pooling in random effect estimates across the levels in
the hierarchy. We elaborate on this benefit in Section~\ref{sec:local-global}.

Despite its appeal, the hierarchical modeling and more general mixed modelling
approaches to recommender systems have long considered infeasible at
commercial scale due to the high computational demands of fitting the model
\citep{Agar08,Naik08}. Recent progress has expanded the scope of application
of these models: \citet{Gao2016arXiv1,Gao2016arXiv2} proposed a moment-based
approach for estimating the parameters of a crossed effects model using a
moment-based approach; \citet{Perr16} proposed a moment-based approach for
fitting a flat hierarchical model; \citet{Tan2018} proposed a kernel-based
approach for fitting a linear flat hierarchical model; and \citet{Zhang2016}
developed a parallelized maximum likelihood fitting algorithm that can exploit
multiple computing cores.

To fully exploit the deeply-nested book hierarchy in a computationally
efficient manner, we will in the sequel develop a moment-based fitting method
for hierarchical models of arbitrary depth. Before doing so, in
Section~\ref{sec:local-global} we elaborate on the benefits of using a
hierarchical model in our context.  Next in
Section~\ref{sec:modelframework} we introduce the details of our model, using
framework suitable for describing general hierarchical models.  Our fitting
method proceeds in two passes. In the first pass, we use the available data to
get initial parameter estimates at the leaves of the tree, and then we
propagate information in these estimates up to the root. In the second pass,
we use the accumulated information to refine the estimates back down from the
root back to the leaves.  We describe these procedures in
Section~\ref{sec:fitting-procedure} and Section~\ref{sec:ebayes-est}, respectively.
After investigating the performance of our method in simulations in
Section~\ref{sec:simulation}, we apply the procedure to our dataset in
Sections~\ref{sec:application} and~\ref{sec:results}. We conclude with
a short discussion in Section~\ref{sec:discussion}.

Our fitting procedure is implemented in the
\texttt{mbest} R package, available at
\url{https://cran.r-project.org/package=mbest}.

\section{Local and global approaches}
\label{sec:local-global}

This hierarchical modeling approach interpolates two extremes: a ``global''
and a ``local'' approach. The global approach would have a single parameter vector
shared by all books, fit by lumping the reviews for all books together. The
local approach would have a different parameter vector for each book, fit
in isolation using only the reviews of the corresponding book. The global
approach corresponds to setting $u_i = 0$ for all books~$i$; the local
approach corresponds to treating $u_i$ as nonrandom.

The appeal of the hierarchical model can be demonstrated by comparing the
performance of a global and a local model. We perform this comparison at
the author level: the global model will have a single parameter vector shared
by all authors; the local model will have a different parameter vector for
each author. Both models use the same set of covariates, described
later in Section~\ref{sec:modelframework}. We fit both models on a training set, then
evaluate on a test set. In the global model the probability that a user
likes an item is described as
\[
    \logit \Pr(y = 1) = \beta^\T x,
\]
where $x$ is the vector of user covariates. The predictions from this model
are the same for all authors. In the local model, the probability that the user
likes book~$i$ is described as
\[
    \logit \Pr(y = 1) = \beta_i^\T x.
\]
This takes a similar form to the global model, but the coefficients $\beta_i$
depend on the book author~$i$.

Table~\ref{table:pooling} shows the test set
misclassification rates for the two fitted models, grouped by the number of
ratings per author. For authors with 100 or fewer ratings, the global model
performs better on average than the local model. Only in the group of authors
with large number of ratings (more than 100) does the local model perform
better.

\begin{table}[h!] 
\centering
\caption{Average Misclassification Rates of Local Author-specific (Error$_{\text{a}}$)
    and Global (Error$_{\text{g}}$) Models. (Standard Deviations in Parentheses.)
    }
\begin{tabular}{|r|r|r|r|} \hline 
    \# Author Ratings  &  Error$_{\text{a}}$ (\%)  & Error$_{\text{g}}$ (\%)& Error$_{\text{a}}$ -Error$_{\text{g}}$ (\%) \\ 
 \hline\hline 
 [10,20]& 38.52 (0.53)& 35.16 (0.52)&  3.36\\
(20,50] & 37.15 (0.42)& 34.91 (0.41)&  2.24\\
(50,100] & 34.76 (0.48)& 34.20 (0.48)&  0.55\\
(100,1000] &  32.15 (0.40)& 32.32 (0.40)& -0.16\\
 \hline 
 \end{tabular} 
 \label{table:pooling} 
 \end{table}

What is needed is an adaptive model that can interpolate between these two
extremes: for authors with abundant data, use the local model; for authors
with little or no data, use the global model; for others, use some combination
of the two models.

The hierarchical model achieves the interpolation between local and global
models automatically. For items~$i$ with abundant data, the posterior
distribution (conditional on the data) for the random effect vector~$u_i$
is concentrated around the local coefficient estimate that uses only the
reviews for item~$i$. For items with no data, the posterior distribution
for $u_i$ is diffuse, with mean zero; the predictions for item~$i$ are
determined mostly by the fixed effect vector~$\beta$ shared globally
by all items. As the number of reviews for item~$i$ increases, the posterior
distribution for $u_i$ and the corresponding predictions for item~$i$
interpolate between these local and global extremes.

The flat item hierarchy shares information across all items.  In our
application, we have a deep hierarchy of books. In
Section~\ref{sec:modelframework}, we will show how to exploit this deep hierarchy
by a using a model with a random effect vector at each node in the hierarchy.
Predictions for the items at the leaves involve the random effects on the
nodes on the path from the root of the hierarchy to the item.  Information
pooling occurs at all levels of the hierarchy, with siblings in a subtree
pooled to estimate the posterior distribution of their random effects.
A hierarchy node with a subtree of abundant data will have an estimated random effect
close to what would come in a fitted model. For other nodes, the estimate will
involve information-pooling across other nodes at the same level in the tree.

\section{Modeling framework}
\label{sec:modelframework}
\label{3l:setup}

For our proxy problem, the goal is to estimate, for a given book and user, the
probability that the user would like the book conditional on the user rating
the book. We have a set of user and context covariates for each review, and a
response~$y$ indicating whether the user liked the book. We also have a
deeply nested hierarchy of books nested under author, sub-subgenere, subgenre,
and genre. In what follows, we describe a model that leverages the book
hierarchy in a way that facilitates information pooling for the estimates
across the levels in the hierarchy.

To introduce the model, we first need to be more precise about what we
mean by a hierarchy in the context of our problem and other similar settings.
For us, a ``depth-$d$ hierarchy'' is a tree where all leaves have depth $d$.
In such a hierarchy, we label the nodes of the tree by unique strings
of natural numbers:
\begin{itemize}
    \item the root of the hierarchy gets labeled by the empty string,
        denoted $\ast$;

    \item the children of the root get labeled by the length-1
        strings $1, 2, \dotsc, M_{\ast}$;

    \item in general, if $i$ is the label of a node, we let $M_i$ denote
        its number of children; we label these children by the strings
        gotten by concatenating the label $i$ with the child
        identifiers: $i1, i2, \dotsc, iM_i$.
\end{itemize}
In this labeling scheme, each node other than the root has a label
that can be represented as $ij$, where $i$ is the node's parent and
$j$ is a natural number in the range $1, 2, \dotsc, M_i$.

For a node $i$, we denote its depth by $|i|$, equal to its distance
from the root; this is also equal to the length of its label. For
any depth $l$ in the range $1, \dotsc, d$, we let $N_l$ denote the
set of nodes with depth $l$. Finally, for node $i$ of depth $l$,
and for $0 \leq k \leq l$, we let $\pi(i, k)$ denote its
ancestor at depth~$k$ in the hierarchy, setting $\pi(i, |i|) = i$.

In the context of our application, the leaves of the hierarchy are authors.
The internal nodes are genres and subgenres. We observe data for each author
(ratings for that author by users); our goal is to relate these ratings
to the user covariates, using the structure of the book hierarchy to
inform our predictions.

In general terms, the hierarchical model supposes that at each leaf node $i
\in N_d$ we observe a response vector $y_i$ of length $n_i$. The behavior of
this response vector is linked to observable covariates through a vector of
nonrandom fixed effects identified with the root of the hierarchy and a set of
random effects identified with the nodes in the hierarchy on the path from the
root to the leaf $i$. Different levels of the hierarchy may use different
sets of user and book features to predict the user's probability of liking
the book. We denote these features by $X_{i0}, \dotsc, X_{id}$, where 
$X_{il}$, a matrix of dimension $n_i \times q_l$, contains the features used
by level-$l$ of the hierarchical model to predict the response vector $y_i$.
To link these features to the response, we posit existence of a fixed
effect vector $\beta$ of dimension $q_0$ identified with the root of the
hierarchy, along with a random effect vector $u_i$ at every other node in
the tree such that $u_i$ has dimension $q_{l}$ when $i$ is at depth $l$. The
distribution of the response $y_i$ is determined by some function of
the linear predictor $\eta_i$, defined as
\begin{equation}\label{eqn:linear-predictor}
    \eta_i = X_{i0} \, \beta + \sum_{l=1}^{d} X_{il} \, u_{\pi(i, l)}.
\end{equation}
This predictor involves the fixed effect vector $\beta$ and the random effects
of all nodes on the path from the leaf $i$ to the root.

In our application, we have a binary response vector $y_i$ with entries
indicating whether the user liked the book. We use the canonical
logistic link, supposing
that for $k = 1, \dotsc, n_i$, this predictor relates to the response as
\begin{equation}\label{eqn:logistic-family}
    \logit \Pr(y_{ik} = 1 \mid u) = \eta_{ik}
\end{equation}
where $u$ without subscript denotes the collection of all random effects.
We further suppose that the components of the vector $y_{i}$ are independent
of each other conditional on $u$.
In an application with a continuous response vector $y_i$ we would instead
typically specify that $y_i$ has independent Gaussian components with mean
and variance given by
\begin{equation}\label{eqn:normal-family}
    \ex(y_{ik} \mid u) = \eta_{ik}, \qquad \var(y_{ik} \mid u) = \phi
\end{equation}
for $k = 1, \dotsc, n_i$ and for some dispersion parameter $\phi$. The
hierarchical model is not limited to these two settings, and in principle
a modeler could specify any link between the linear predictor $\eta_i$
and the mean of the response~$y_i$.

To endow our model with a mechanism that allows borrowing strength across
similar items in the hierarchy, we model the $d$ populations of random effects
at the levels of the hierarchy. We treat these populations as independent.
For the population of level-$l$ random effects, $l=1, \dotsc, d$,
we suppose that each item $u_i$ is an independent draw from a multivariate
normal distribution with mean-zero and covariance matrix
$\Sigma_l$ for some $q_l \times q_l$ covariance matrix~$\Sigma_l$:
\begin{equation}\label{eqn:random-effect}
    u_i \sim \Normal(0, \Sigma_l) \qquad \text{for $i \in N_l$.}
\end{equation}
We further suppose that all random effects $u$ are independent of each other.

In the sequel, we discuss estimation for the depth-$d$ hierarchical model. That
estimation procedes in two stages: first, estimate the model parameters $\beta$
and $\Sigma_1, \dotsc, \Sigma_d$. Next, use the model parameters to get
empirical Bayes estimates of the random effects $\{ \hat u_i \}$. The
empirical Bayes estimation procedure is the part of the model estimation that
leverages information across different levels of the hierarchy. Our estimates
of the covariance matrices $\Sigma_1, \dotsc, \Sigma_d$ allow
us to impute components of particular random effects vectors when we only
have information about a subset of their components.

\section{Fitting procedure}
\label{sec:fitting-procedure}

Frequentist hierarchical models like the one described in the previous section
often get fit via maximum likelihood~\citep{lme4}. However, these fitting
algorithms can be prohibitively slow for large data sets like those that
appear in commercial-scale settings~\citep{Agar08,Naik08}.  \citet{Perr16} got
over the computational hurdle in a depth-1 hierarchical model by using a
moment-based estimation procedure, adapted from an earlier procedure due to
\citet{Coch37}. Here, we will extend \citepos{Perr16} procedure to handle
hierarchy of arbitrary depth.

Throughout the section we will assume that the model described
in~\eqref{eqn:linear-predictor} and~\eqref{eqn:random-effect} is in force.
For the response data $y_i$ and the leaf nodes $i \in N_d$ we will allow for
both the logistic regression case from~\eqref{eqn:logistic-family}, the normal
response case from~\eqref{eqn:normal-family}. Our estimators extend naturally
to any generalized linear model at the leaves with shared dispersion
parameter~$\phi$.

The estimation procedure is easiest to describe if we reparametrize. To do so,
for any node $i$, let $b_i$ denote the vector of fixed and random effects on
the path from the root up to and including $i$. Specifically, set $b_{\ast} =
\beta$ and for node $ij$ with parent $i$ define recursively
\[
    b_{ij} = (b_i, u_{ij}).
\]
For depth $l = 1, \dotsc, d$, let $p_l$ be the
total number of fixed and random effects up to and including depth $l$:
\[
    p_l = q_0 + q_1 + \dotso + q_l;
\]
if $i \in N_l$, then $b_i$ has $p_l$ components.

Now, for leaf node $i \in N_d$ define the matrix
gotten by
concatenating the columns of feature matrices by
$X_{i} = [ X_{i0}\ X_{i1} \cdots \ X_{id} ]$, so that
$X_{i}$ has dimension
$n_{i} \times p_d$.  In this reparametrized form, the linear predictor at
leaf node~$i$ is
\begin{equation}\label{eqn:linear-predictor-alt}
    \eta_{i} = X_{i} \, b_{i}.
\end{equation}
This form makes the hierarchical model look somewhat like a standard
generalized linear model, but the effect vector $b_i$ includes both
nonrandom and random components: the fixed effect vector $\beta$ and the
random effects $u_{\pi(i, l)}$ on the path from the root to the leaf~$i$.

The estimation procedure for the hierarchical model is defined by repeatedly
pruning the tree by reducing the leaves to a set of estimates at their
parents. The high level description of the procedure is as follows:

\begin{enumerate}
    \item Produce estimates $\hat b_{i}$ of $b_{i}$
        at each leaf node $i \in N_d$. Set $l = d$.

    \item We have at hand estimates $\hat b_{ij}$ of
        $b_{ij} = (b_i, u_{ij})$ for each node $ij \in
        N_{l}$.  For each $i \in N_{l-1}$, combine the child estimates,
        $\hat b_{ij}$ for $j = 1, \dots, M_i$, to produce an
        estimate $\hat b_{i}$ of $b_i$ and an estimate
        $\hat \Sigma_{li}$ of $\Sigma_l$.

    \item Combine estimates $\hat \Sigma_{li}$ for $i \in N_{l-1}$ to produce
        a final estimate $\bar \Sigma_l$.

    \item If $l = 1$, set
        $\bar \beta = \hat b_{\ast}$ to be the final estimate of
        the fixed effects and stop. Otherwise, go to Step~2 with the level $l$
        decreased to $l - 1$.
\end{enumerate}

\noindent
In settings with a dispersion parameter $\phi$, we handle this parameter
analogously to $\Sigma_d$.

When the fitting procedure terminates we will have final estimates $\bar \beta$
and $\bar \Sigma_1, \dotsc, \bar \Sigma_d$ of the fixed effects and
the random effect covariance matrices.
The rest of this section is devoted to detailing the individual steps
of the fitting procedure.

\subsection{Step 1: Estimate parameters at the leaves}

The first step in the estimation procedure is to use the data $y_{i}$ at each
leaf $i \in N_d$ to produce an estimate $\hat b_{i}$ of $b_{i}$, the
vector of fixed and random effects on the path from the root to the leaf. 
Recall that $\eta_i = X_i b_i$.
We will explicitly handle cases where the combined predictor matrix $X_{i}$
is rank-degenerate.  We will only require that, conditional on $b_{i}$, the
estimate $\hat b_{i}$ has negligible bias outside the null space of $X_{i}$
and is approximately normally distributed with known covariance matrix.

First we handle the normal model~\eqref{eqn:normal-family}, where for
$k = 1, \dotsc, n_i$ the components of the response satisfy
$y_{ik} = \eta_{ik} + \varepsilon_{ik}$ for a mean-zero Gaussian error
vector $\varepsilon_i$ with independent components
$\varepsilon_{ik}$ for $k = 1, \dotsc, n_i$ having unknown variance $\phi$. In this case
we set
\[
    \hat b_{i} = (X_{i}^\T X_{i})^{\pinv} X_{i}^\T y_{i},
\]
where $\pinv$ denotes pseudo-inverse. When $X_{i}$ has full rank,
the estimate $\hat b_{i}$ is the unique least-squares estimate of $b_{i}$;
otherwise, the least-squares estimate is not unique and $\hat b_{i}$ is
one of the vectors minimizing the squared Euclidean norm
$\| y_{i} - X_{i} \hat b_{i} \|^2$.

To define the estimate of the dispersion parameter $\phi$, we let $r_{i}$ denote the rank of
$X_{i}$. When $r_{i} < n_{i}$ we set
\[
    \hat \phi_{i} = \frac{1}{n_{i} - r_{i}} \| y_{i} - X_{i} \hat b_{i} \|^2;
\]
otherwise, we set $\hat \phi_{i} = 0$.
We combine the estimates $\hat \phi_i$ across all the leaves to get a single
estimate for the dispersion parameter:
\[
    \bar \phi = \frac{\sum_{i \in N_d} (n_i - r_i) \hat \phi_i}
                     {\sum_{i \in N_d} (n_i - r_i)}.
\]

Next we derive the properties of the estimate $\hat b_i$. First, let
$X_{i} = U_i D_i V_i^\T$ denote a compact singular value decomposition where
$D_i$ is a diagonal matrix of dimension $r_i \times r_i$ with positive
diagonal entries. Then
\begin{align*}
    \hat b_i
    &= V_i D_i^{-1} U_i y_i \\
    &= V_i V_i^\T b_i + V_i D_i^{-1} U_i^\T \varepsilon_i.
\end{align*}
Thus,
\[
    D_i V_i^\T (\hat b_i - b_i) = e_i,
\]
where $e_i = U_i^\T \varepsilon_i$ is a mean-zero Gaussian random
vector of $r_i$ independent components, each with variance $\phi$. If we set
$Z_i = \bar \phi^{-1/2} D_i V_i^\T$, then the quantity
$Z_i (\hat b_i - b_i)$ is approximately mean-zero normal with identity
covariance matrix.

For the logistic regression model~\eqref{eqn:logistic-family}, we proceed
analogously, but we use the Firth's biased-reduced estimator \citep{Firt93} in
place of the least squares estimator for $b_i$. This is a refinement of the
maximum likelihood estimator that is well-defined even when the responses are
perfectly separated by a linear combination of the predictors.  In cases where
$X_i$ is rank-degenerate, there are multiple such estimators; we arbitrarily
take $\hat b_i$ to be one of them. The properties of the estimator are like
those of the maximum likelihood: as the sample size $n_i$ increases,
the estimator is asymptotically unbiased with covariance equal to the
inverse information matrix. In the case of rank-deficient feature matrix
$X_i$, the information matrix takes the form $I(b_i) = V_i D_i(b_i) V_i^\T$
where $V_i$ is the matrix of right singular vectors of $X_i$. In this
case if we set $Z_i = D_i(\hat b_i) V_i^\T$, then
$Z_i (\hat b_i - b_i)$ is approximately mean-zero normal with identity
covariance matrix.

In both the normal and the logistic regression case, we can find an estimator
$\hat b_i$ and a matrix $Z_i$ with full column rank $r_i$ such that
conditional on $b_i$, the quantity $Z_i^\T (\hat b_i - b_i)$ is
approximately normal with identity covariance.  In the logistic regression
case, the quality of the normal approximation depends on the sample size $n_i$
being large.

\subsection{Step 2: Combine the estimates at level $l$}

We now suppose that for some level $l$, for each node $ij \in N_l$
we have a matrix $Z_{ij}$ of full row rank $r_{ij}$, 
such that conditional on $b_{ij}$,
\begin{align*}
    \ex  \{ Z_{ij} (\h b_{ij} - b_{ij}) \mid b_{ij} \} = 0, \quad 
    \cov \{ Z_{ij} (\h b_{ij} - b_{ij}) \mid b_{ij} \} = I.
\end{align*}
For linear models, these two conditions hold exactly; in nonlinear models
these will only hold approximately, with the quality of the approximation
depending on the size of the sample used to estimate $\h b_{ij}$.
We will show how to combine estimates $\h b_{i1}, \dotsc, \h
b_{iM_i}$ to get an estimate $\h b_i$ of $b_i$ and an estimate $\h\Sigma_{il}$
of $\Sigma_{l}$.

Recall that $b_{ij} = (b_i, u_{ij})$ for each node $ij\in N_l$, where $u_{ij}$ is the random 
effects on level $l$, and $b_i, u_{ij}$ are of length $p_{l-1},q_l$ respectively.
Let $Z_{ij} = U_{ij} D_{ij} V_{ij}^T$ be a compact singular value decomposition.
When the context is clear, for simplicity we denote $V_{ij1}$ as the first $p_{l-1}$ rows
of $V_{ij}$, and denote $V_{ij2}$ as the last $q_l$ rows of $V_{ij}$.

We have the following (unconditional) moment equations:
\begin{align}
    \label{eq:2lmoment_1}
    \ex(V_{ij}^T \h b_{ij} ) &= V_{ij}^T (b_i ,0) = V_{ij1}^T b_i, \\
    \label{eq:2lmoment_2}
    \text{cov}(V_{ij}^T \h b_{ij} ) &=  D_{ij} ^{-2} +
V_{ij}^T \left[\begin{array}{c c} 0 & 0 \\ 0 & \Sigma_{l} \end{array}\right] V_{ij}
    =  D_{ij} ^{-2} + V_{ij2}^T \Sigma_{l} V_{ij2}.
\end{align}

The moment equations~\eqref{eq:2lmoment_1}
and~\eqref{eq:2lmoment_2} hold for any node $ij\in N_l$, therefore
by standard moment matching method, we want to take empirical mean
of terms on the left hand side and set parameters on the right hand side
to match it. However, we cannot do this right away, since the dimension of
$V_{ij}^T$, or equivalently the rank $r_{ij}$, may vary by nodes $ij\in N_l$. 
To overcome this, we augment the moment equations to have same dimension across nodes $ij \in N_l$. 
In particular, with any choice of symmetric positive-definite matrix $W_{ij}$ for each node $ij\in N_l$, 
we have 
\begin{align}
    \label{eq:2lwmoment_1}
    \ex(V_{ij1} W_{ij} V_{ij}^T \h b_{ij} ) &= V_{ij1} W_{ij} V_{ij1}^T b_i, \\
    \label{eq:2lwmoment_2}
    \cov(V_{ij2} W_{ij} V_{ij}^T \h b_{ij} ) 
    & = V_{ij2} W_{ij} ( D_{ij} ^{-2} + 
V_{ij2}^T \Sigma_{l} V_{ij2} )W_{ij} V_{ij2}^T.
\end{align}
We use the semi-weighted scheme for choosing $W_{ij}$ as described by
\citep{Perr16}.

Now, the moment equations have consistent dimension across all nodes:
for every node $ij\in N_l$, equation~\eqref{eq:2lwmoment_1} has dimension $p_{l-1}\times 1$,
and \eqref{eq:2lwmoment_2} has dimension $q_l\times q_l$.
Based on equation~\eqref{eq:2lwmoment_1}, we define
the moment-based estimator $\h b_i$ as
\[
    \h b_i = \Omega_i ^\dagger \sum_{j = 1}^{M_i} V_{ij1} W_{ij} V_{ij}^T \h b_{ij} ,
    \quad \Omega_i = \sum_{j = 1}^{M_i} V_{ij1} W_{ij} V_{ij1}^T,
\]
where $\dagger$ denotes pseudo-inverse.
Based on equation~\eqref{eq:2lwmoment_2}, 
the moment-based estimator $\h \Sigma_{il}$ should satisfy
\begin{multline*}
    \sum_{j = 1}^{M_i} (V_{ij2} W_{ij} V_{ij}^T \h b_{ij} -V_{ij2} W_{ij} V_{ij1}^T b_i)
    (V_{ij2} W_{ij} V_{ij}^T \h b_{ij} -V_{ij2} W_{ij} V_{ij1}^T b_i)^T \\
    \\
    = \sum_{j = 1}^{M_i} V_{ij2} W_{ij}  D_{ij} ^{-2} W_{ij} V_{ij2}^T 
    +  \sum_{j = 1}^{M_i}  V_{ij2} W_{ij} V_{ij2}^T \h\Sigma_{il} V_{ij2} W_{ij} V_{ij2}^T.
\end{multline*}
In practice, we do not have access to the true $b_i$
to compute $\h \Sigma_{il}$ in the above equation,
instead we use the empirical estimate $\h b_i$.
If the result $\h\Sigma_{il}$ is not positive semidefinite, we project
it onto the cone of positive semidefinite matrices and obtain the final estimate.

Let $\Omega_i = V_i D_i V_i^T$
denote the eigendecomposition of the positive semidefinite matrix $\Omega_i$. 
Let $\Omega_i^{1/2}$ denote the symmetric square root of $\Omega_i$.
\citepos{Perr16} results imply that, subject to assumptions on the sample
size, conditional on $b_i$, the quantity $\Omega_i^{1/2} (\h b_i - b_i)$ is
approximately normally distributed with
\begin{align*}
    \ex \{  \Omega_i^{{1}/{2}} (\h b_{i} - b_{i}) \mid b_{i} \} =0, \quad 
    \cov \{ \Omega_i^{{1}/{2}} (\h b_{i} - b_{i}) \mid b_{i}  \} \approx V_i V_i^T.
\end{align*}
The error in the approximation tends to zero as the number of child nodes
$M_i$ increases.
In addition, since
$V_{i}^T \Omega_i^{{1}/{2}} = V_i^T V_i D_i^{{1}/{2}} V_i^T = D_i^{{1}/{2}} V_i^T$.
Thus we can rewrite the above results as 
\begin{align}\label{eq:new_construct}
    \ex \{  D_{i}^{{1}/{2}} V_i^T (\h b_{i} - b_{i}) \mid b_{i} \} =0, \quad 
    \cov \{ D_{i}^{{1}/{2}} V_i^T (\h b_{i} - b_{i}) \mid b_{i}  \} \approx I.
\end{align}
\citet{Perr16} details the precise assumptions
required for these results along with the quality of the approximations.

\subsection{Step 3: Combine the level-$l$ covariance estimates}

At the end of Step~2 in the procedure we have an estimates $\h\Sigma_{il}$ of
$\Sigma_l$ for each $i \in N_{l-1}$. In Step~3, we combine these estimates
to produce a final estimate $\bar \Sigma_l$ by taking a weighted average:
of 
$\h\Sigma_{il}$ over all nodes $i\in N_{l-1}$,
\begin{align*}
    \bar  \Sigma_{l}  =  \frac{\sum_{i \in N_{l-1}} M_i \h \Sigma_{il}}
    {\sum_{i \in N_{l-1}} M_i}.
\end{align*}
Nodes with higher numbers of children $M_i$ get more weights.

\subsection{Step 4: Recurse or Stop}

If we are at the root of the tree, so that $l = 0$ and we have an
estimate $\hat b_{\ast}$, then we terminate the estimation procedure
by setting our final estimate of the fixed effects to $\bar \beta = \hat
b_{\ast}$.  Otherwise we decrement $l$ to $l - 1$,
and go to Step~2 with $Z_i=D_i^{{1}/{2}}V_i^T$ that are used in equation~\eqref{eq:new_construct}.

\section{Empirical Bayes random effect estimates}
\label{sec:ebayes-est}

At the end of the fitting procedure described in
Sec.~\ref{sec:fitting-procedure} we have estimate $\bar \beta$ of
the fixed effect vector and estimates $\bar \Sigma_1, \dotsc, \bar \Sigma_d$
of the random effect covariance matrices. We also have at each node~$i$
in the hierarchy a preliminary estimate $\hat b_i$ of $b_i$, the fixed and
random effects on the path from the root to node~$i$. These preliminary
estimates do not share information across the hierarchy; estimate $\hat b_i$
is determined only from the data at the leaves descending from $i$. We
can improve the estimates by replacing each $\hat b_i$ with an empirical Bayes
estimate $\bar b_i$ that pools information across the hierarchy.

The information-pooling algorithm works top-down from the root. It starts
by setting $\bar b_\ast = \bar \beta$. Then at depth-1 nodes $j \in N_1$,
the procedure uses $\bar b_\ast$ and $\bar \Sigma_1$ together with $\hat b_j$
to get a refined estimate $\bar b_j$. This process repeats, level by level,
until we get refined estimates at the leaves.

The full procedure is as follows:
\begin{enumerate}
    \item Set $\bar b_\ast = \bar \beta$ and set $l = 0$.

    \item If $l = d$, stop.

    \item For each node $i \in N_l$ we have a refined estimate $\bar b_i$.
        For each child $ij$ for $j = 1, \dotsc, M_i$, we have a
        preliminary estimate $\hat b_{ij}$. Use $\bar b_i$ together
        with $\hat b_{ij}$ and $\bar \Sigma_{l+1}$ to produce a refined
        estimate $\bar b_{ij}$.

    \item Increment $l$ to $l + 1$ and go to Step~2.
\end{enumerate}

\noindent
After applying this procedure, we have a refined estimate $\bar b_i$ at
each node in the tree. The estimates at each leaf~$i$ can be used to
make refined estimates of the linear predictors ($\bar \eta_i = X_i \bar
b_i)$ or they can be used to make predictions for new data.

To complete the description of the procedure we need to explain Step~3 in
more detail. In this step we have at our disposal $\bar b_i$, $\bar
\Sigma_{l+1}$, and $\h b_{ij}$ for node $i\in N_l$ and its children.
Further, we have a matrix $Z_{ij}$ of
full column rank $r_{ij}$ such that $Z_{ij} (\hat b_{ij} - b_{ij})$
is approximately distributed as a multivariate normal with identity
covariance matrix.

By definition, $b_{ij} = (b_i, u_{ij})$ where $u_{ij}$ is the random 
effect vector for node $ij\in N_{l+1}$, and $b_i, u_{ij}$ are of length $p_{l},q_{l+1}$ respectively.
We denote $Z_{ij1}$ as the first $p_{l}$ columns
of $Z_{ij}$, and denote $Z_{ij2}$ as the last $q_{l+1}$ columns of $Z_{ij}$. 
Conditional on $b_i$, we have the following (approximate) Bayesian linear regression model:
for any node $j\in N_{l+1}$,
\begin{align*}
    u_{ij} &\sim \Normal(0, \Sigma_{l+1}),\\
    Z_{ij} \h b_{ij} & = Z_{ij1} b_i + Z_{ij2} u_{ij} + e_{ij},\\
    e_{ij} & \sim  \Normal(0, I).
\end{align*}
The empirical Bayes estimate $\hat u_{ij}$ is an estimate of the
posterior mean of $u_{ij}$ conditional on the observed data $Z_{ij} \hat b_{ij}$,
gotten by using plug-in estimates $\bar b_i$ and $\hat \Sigma_{l+1}$ for
$b_i$ and $\Sigma_{l+1}$.

To derive the posterior distribution 
define $Y_{ij} = Z_{ij} \h b_{ij}  - Z_{ij1} b_i $, noting that the
conditional distribution $u_{ij} \mid Z_{ij} \h b_{ij}, b_i$ is the
same as that of $u_{ij} \mid Y_{ij}, b_i$.
By Bayes rule, then, the posterior density of $u_{ij}$ satisfies
\begin{align*}
    p(u_{ij} \mid Y_{ij},b_i)
     & \propto p(Y_{ij} \mid u_{ij},b_i) \, p(u_{ij}) \\
    & \propto \exp\{-\tfrac{1}{2} 
    (Y_{ij} - Z_{ij2} u_{ij})^T 
    (Y_{ij} - Z_{ij2} u_{ij})
    -\tfrac{1}{2} u_{ij}^\T \Sigma_{l+1}^{-1} u_{ij} \} \\
    & \propto  \exp[-\tfrac{1}{2} \{
    u_{ij}^\T (Z_{ij2}^\T Z_{ij2} + \Sigma_{l+1}^{-1}) u_{ij} 
    - 2 Y_{ij}^\T Z_{ij2} u_{ij} 
     \}].
\end{align*}
The posterior distribution, then, is that of a multivariate Gaussian
with expected value given by
\begin{align*}
    \ex(u_{ij}\mid Z_{ij} \h b_{ij}, b_i)
     = (Z_{ij2}^T Z_{ij2} + \Sigma_{l+1}^{-1})^{-1} Z_{ij2}^\T (Z_{ij} \h b_{ij} - Z_{ij1} b_i).
\end{align*}
The empirical Bayes estimate of $u_{ij}$ comes from using this expression in
conjunction with plug-in estimates for $\Sigma_{l+1}$ and $b_i$:
\[
    \hat u_{ij}
     = (Z_{ij2}^T Z_{ij2} + \hat \Sigma_{l+1}^{-1})^{-1} Z_{ij2}^\T (Z_{ij} \h b_{ij} - Z_{ij1} \bar b_i).
\]
The refined estimate of $b_{ij}$, then, is $\bar b_{ij} = (\bar b_i, \hat u_{ij})$.

\section{Simulation}
\label{sec:simulation}

To evaluate our proposed estimation method, we compare its performance with
three other procedures:
\begin{itemize}
\item{\texttt{glmer}}, a maximum likelihood procedure, implemented
    as part of the \texttt{lme4} R~package \citep{lme4}.

\item{\texttt{glmer.split}}, a data-splitting estimation procedure, which
    randomly splits the data set into 10 subsets, computes estimates on each
    of them separately using \texttt{glmer}, and then combine the
    estimates by averaging them. We implemented the
    procedure ourselves in~R; the algorithm is based on procedures
    proposed by~\citet{Huan05}, \citet{Gebr12}, and \citet{Scot13B}.

\item{\texttt{sgd}}, which uses stochastic gradient descent to maximize a
    regularized version of the $h$-likelihood. We implemented the procedure in
    a combination of~C and ~R; the algorithm is based on procedures
    proposed by~\citet{Kore09b} and \citet{Dror11}.  We choose the
    regularization parameters by cross-validation.
\end{itemize}
In evaluating the methods, we look at both the quality of their estimates
and the time it takes to compute them. We do not include the tuning parameter
cross-validation time in the timing results.

We perform two sets of simulations: one for a two-level logistic regression
model, and one for a two-level linear regression model. The setup and results
for both simulations are similar, so we only include the logistic regression
results here.  Appendix~\ref{app:twolevel-linear} contains the linear
regression results.

Following the notation in Section~\ref{3l:setup}, we set the number of groups
on the first level to $|N_1|=50$, and number of groups on second level (the leaves)
to $|N_2|=500$.  We simulate $N$ samples with $N$ ranging from $1000$ to
$100000$.  We set the dimensions of fixed and random effect vectors to
$q_0=q_1=q_2=5$.  For each value of $N$ we draw $20$ replicates according to
the following procedure.

For each replicate, we draw a $q_0$-dimensional fixed effect vector $\beta$ with components 
$\beta_{k},k=1,\dots,q_0$ drawn independently from a heavy-tailed
student's $t$-distribution with 4~degrees of freedom. 
We draw random effect covariance matrices $\Sigma_{1}$ and $\Sigma_{2}$ independently from an 
inverse Wishart distribution with shape $I$ and 10 degrees of freedom, scaled by $0.1$.

We allocate the~$N$ samples to the $50$ groups and $500$ subgroups, in a way
that approximates the highly skewed
hierarchies in the Book Crossing dataset. In each replicate, we first draw sampling rates
$\lambda_1,\dots,\lambda_{500}$ from a Pareto distribution, with scale and 
shape parameters set to 1. Then, we allocate the $N$ samples to the $500$ leaf
nodes by drawing from a multinomial distribution with probability vector
$(\lambda_1,\dotsc, \lambda_{500})/\sum_{i=1}^{500} \lambda_i$. 
Similarly we allocate the $500$ leaf nodes to $50$ groups 
using the same Pareto distribution and sampling scheme. 
 
For every group node in the first level of the hierarchy, $i\in N_1$, 
we draw a $q_1$-dimensional random effect vector $u_{i}$ 
from multivariate Gaussian with mean zero and covariance matrix $\Sigma_1$; 
for every leaf node $i \in N_2$, we draw a $q_2$-dimensional random effect vector 
$u_{i}$ from multivariate Gaussian with mean zero and covariance $\Sigma_2$. 
Then we randomly draw fixed effect predictor vectors $x_k$ for sample point $k=1,\dots,N$, 
with independent elements taking values~$+1$ and~$-1$ with probability
${1}/{2}$ each.
We use the same procedure to randomly draw random effect predictors $z_k$ for every sample point $k$, 
and let the two levels of the hierarchy share the same random effect predictors. 
Finally, for every sample $k$ in leaf node $ij \in N_2$, we draw response $y_k$ as 
Bernoulli with success probability
\[
    \mu_k=\logit^{-1}\{ x_k^\T\beta+z_k^\T(u_{i}+u_{ij}) \}.
\]

To evaluate the quality of the estimators, we use the following loss functions:
\begin{itemize}
\item{Fixed Effect Loss}: $\Vert \beta-\h{\beta}\Vert ^{2}$;
\item{Random Effect Level-$l$ Covariance Loss}:
$ \text{tr}\{ (\h{\Sigma}_l\Sigma_l^{-1}-I)^{2}\}$;
\item{Random Effect Level-$l$ Loss}:
    $|N_l|^{-1} \sum_{i\in N_l}\Vert \Sigma_l^{-1/2}(u_{i}-\h u_{i})\Vert ^2$
\item{Prediction Loss}:
\begin{equation*}
    N^{-1}\sum_{k=1}^{N}\mu_{k}\log\frac{\mu_k}{\h{\mu}_k}+
    (1-\mu_{k})\log \frac{1-\mu_{k}}{1-\h{\mu}_k}
\end{equation*}
        where $\hat \mu_k=\logit^{-1}\{ x_k^\T \hat \beta+ z_k^\T(\hat u_{i}+ \hat u_{ij})\}$
for sample~$k$ in leaf~$ij$.
\end{itemize}
We also measure the overall computation time for each,
excluding the
cross-validation time for tuning parameter selection.

\begin{figure}
\centering
\includegraphics{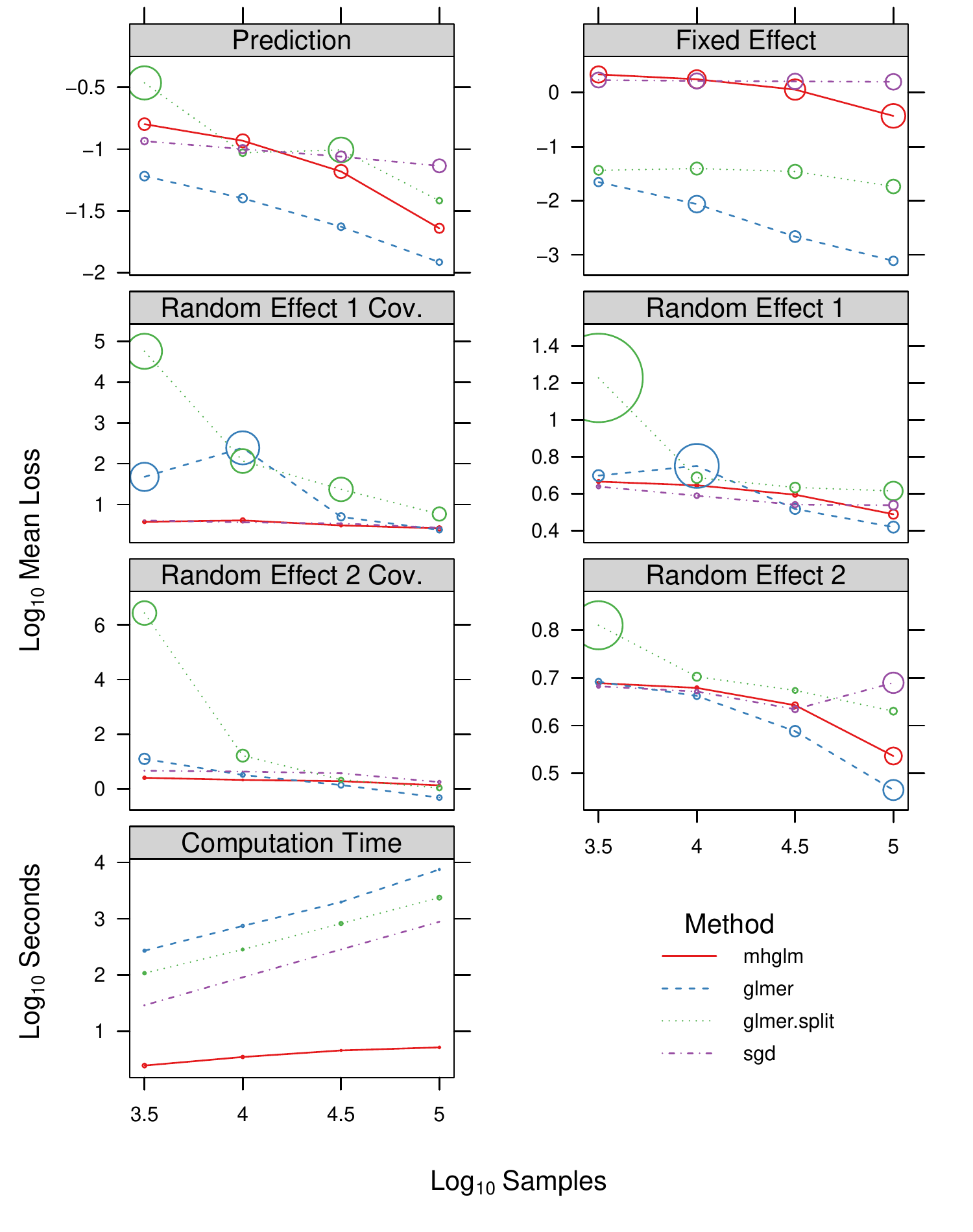}
\caption{Performance for the multilevel logistic regression model.
 radii indicate one standard error along $y$-axis (absent when smaller than line width)}
\label{fig:sim-logistic}
\end{figure}

We compare our method (\texttt{mhglm}) with the three other methods described
above: \texttt{glmer}, \texttt{glmer.split}, and \texttt{sgd}.
Figure~\ref{fig:sim-logistic} shows the mean performance for each method,
averaged over $20$ replicates, with circle radii indicating standard errors
along the vertical axes. For moderate to large sample sizes, there is a
noticeable difference between the proposed method and other maximum likelihood
based estimators. However the proposed method still appears to be consistent,
in the sense that its estimators improve with more samples. In terms of
prediction loss, our proposed method outperformed both $\texttt{sgd}$ and
$\texttt{glmer.split}$ and is only slightly worse than $\texttt{glmer}$.

The bottom panel compares the computation time for all methods. For large
sample sizes, our proposed method is much faster than the other procedures, by
factor ranging from $100$ to $1000$, and the factor appears to grow
exponentially as sample sizes increase.

In the context of this simulation, our proposed method is able to trade off a
modest loss in prediction performance for a dramatic decrease in computation
time. We can see that our proposed procedure will scale well to our book
recommendation context and to commercial recommendation settings generally.

\section{Application}
\label{sec:application}

Having developed an estimation procedure for deeply-nested hierarchical
models in Secs.~\ref{sec:fitting-procedure} and~\ref{sec:ebayes-est}, and
having established its suitability in Sec.~\ref{sec:simulation}, we now return
to our main application, fitting a model to data that allows us to predict
whether or not a user would like a book if he or she had rated it.

Recall from Sec.~\ref{sec:intro} that our dataset consists of two
parts: a set of user ratings of books, and a hierarchy of these books.  We
treat each rating as an observation containing book and user identifiers along
with a numerical score between 1 and 10. To smooth differences between
user-specific rating scales, we binarize the ratings, treating numerical
scores of 8 or above as ``positive'' and ratings below this as ``negative''.
We will model these binarized ratings using a hierarchical logistic regression
model.  We use the user demographic features together with the rating context
to construct candidate predictors in the model, linked to the response through
fixed and random effects. We use a subset of the book hierarchy for the
structure of the hierarchical model.

\subsection{Candidate predictors}

The first set of candidate predictors are converted from user demographic
data.  We bin users' ages into 5 groups: (0,26], (26,32], (32,38], (38,47] and
(47,101], where each group has approximately same number of ratings. Each age
group is represented by one categorical variable.  If age is missing, then all
five indicators are zero.  We aggregate the geographic feature into 6 groups
by continent: North America, Europe, Oceania, Asia, South America, and Africa.
Similarly, each group is represented by one categorical variable.  We have a
total of 11 demographic predictors.

Our second set of candidate predictors is the time-varying predictors defined as
functions of past user behavior. The first predictor \texttt{prev} is a
user-specific binary indicator of whether the user's previous rating was
positive. This is designed to capture a user's propensity to make positive
ratings. The
second predictor, \texttt{dist}, is user-category specific, computed as the
smoothed log proportion of past ratings that the user gives in each category.
This is designed to capture users' tendency to give ratings to each category,
revealing his or her relative preference among all categories.
Table~\ref{table:predictors} gives detailed descriptions of all the
predictors. 

\begin{table}
\begin{center}
\caption{Predictor associated with one observation from user$_i$ on book$_j$. }
\label{table:predictors}
\begin{tabular}[t]{  L{2.1cm} | L{9.8cm} } 
\hline
Predictor & Description  \\
\hline
 Age$_i$ & User-specific features: a 5-component indicator vector for age range 
 (0,26],(26,32],(32,38],(38,47],(47,101]  \\ 
 \hline
 Geographic$_i$ &  User-specific features: a 6-component indicator vector for continent Africa, Asia, Europe, 
 North America, Oceania, South America\\
 \hline
 Previous$_i$ & User-specific feature: a smoothed estimate of the log of proportion of positive 
 ratings from user$_i$:
  $log(p_i + 1)/(n_i+2)$, where $p_i$ and $n_i$ are the number of positive ratings ($\ge 8$)
 and total number of ratings from user$_i$.\\ 
 \hline
 Distribution$_{ij}$ & User-book-specific feature: a smoothed estimate of the log of 
 proportion of ratings in book$_j$'s genre from user$_i$:
 $\log(k_{ij}+1)/(n_i+m)$, where $k_{ij}$ is number of ratings user$_i$ gives in book$_j$'s genre;
 $n_i$ is total number of ratings from user$_i$; and $m$ is total number of genres.  \\
 \hline
\end{tabular}
\end{center}
\end{table}

\subsection{Model selection}
\label{subsec:modelselection}

We perform two forms of model selection. First, we need to choose which parts
of the book hierarchy to use. Second, we need to choose which predictors to
use. To carry out the model selection, we randomly partition the data into
80/10/10 percent chunks, for training/development/testing sets. We train
various models on the training set, select a model with best performance on
the development set, and finally compare the chosen model with other fitting
methods on the testing set.  In data processing, we only use training data to
construct the new features.

We use all predictors for fixed effects, and we can fit these reliably given
the large volume of data. However, we do not use all of these predictors on
all random effects levels. Fitting the random effects is much more difficult,
because it relies on ratings specific to the particular position in the
hierarchy. The population structure of the random effects mitigates against
some of this data sparsity, but there will still be situations where
using coarser hierarchy makes the model less susceptible to
over-fitting. To guard against overfitting in the random effects terms of the
model, we perform model selection by using out-of-sample prediction
performance on the development set.  To choose the specific subset of the
predictors to use as random effects, we fit all possible combinations at all
levels of the model, selecting the model with the lowest misclassification
rate on the development set.

We start with a depth-1 model. 
Here we fit depth-1 models with all five possible grouping factors: genre,
subgenre, sub-subgenre, author, book.  Table~\ref{table:onelevel_all} lists
the best one-level model using each grouping factor. We sort the performance
by misclassification error on development dataset. We see that using author as
the grouping factor, and demographic information as the random effect features
gives the best prediction performance on development set.  Note that we did not
get additional performance improvement by using a book-specific random effect
model, which suggest that we could potentially over-fit the data by using too
many groups.

\begin{table}[h!]
    \centering
    \caption{Best performing model for all choices of grouping factor for one-level model.
    The standard deviations of the listed model errors are below $0.004$}
    \begin{tabular}{| c | c | r |}
        \hline
        Group & Features & Error\\
        \hline\hline
        author&\texttt{geo}&0.3212\\
        book&\texttt{age}, \texttt{geo}&0.3235\\
        sub-subgenre&\texttt{prev}, \texttt{dist}, \texttt{age}, \texttt{geo}&0.3282\\
        subgenre&\texttt{prev}&0.3288\\
        genre&\texttt{1}&0.3291\\
        \hline
    \end{tabular}
    \label{table:onelevel_all}
\end{table}

To further take advantage of the five nested hierarchies, we also consider
depth-2 models.
Using the \texttt{lme4} modeling notation, we fit all models of the following
form:
\[
    y \sim \texttt{age} + \texttt{geo} + \texttt{prev} + \texttt{dist} 
    + (X_1 | g_1) + (X_2 | g_1:g_2).
\]
where grouping level $g_2$ is nested under $g_1$, and $X_1$ and $X_2$
are predictor matrices with columns taken from the candidate
predictors. The notation indicates that the model has fixed effects
corresponding to an intercept and predictors \texttt{age}, \texttt{geo}, \texttt{pref},
and \texttt{dist}, random effect predictors $X_1$ at the first level,
and random effect predictors $X_2$ at the second level.

\begin{table}[h!]
    \centering
    \caption{Best performing model for all choices of grouping factors for two-level model.
    The standard deviations of the listed model prediction errors are below $0.004$}
    \begin{tabular}{| c | c | c | c | r |}
        \hline
        $g_1$ & $g_2$ & $X_1$ & $X_2$& Error\\
        \hline\hline
        subgenre&	author&	\texttt{dist}, \texttt{age}&	\texttt{age}, \texttt{geo}&	0.3177\\
        sub-subgenre&	author&	\texttt{age}&	\texttt{dist}, \texttt{geo}&	0.3184\\
        genre&	author&	\texttt{dist}, \texttt{geo}&	\texttt{age}, \texttt{geo}&	0.3189\\
        author&	book&	\texttt{geo}&	\texttt{dist}&	0.3210\\
        sub-subgenre&	book&	\texttt{dist}, \texttt{age}, \texttt{geo}&	\texttt{geo}&	0.3212\\
        subgenre&	book&	\texttt{prev}, \texttt{dist}, \texttt{age}&	\texttt{age}, \texttt{geo}&	0.3218\\
        genre&	book&	\texttt{age}, \texttt{geo}&	\texttt{age}&	0.3226\\
        subgenre&	sub-subgenre&	\texttt{age}, \texttt{geo}&	\texttt{dist}&	0.3266\\
        genre&	sub-subgenre&	\texttt{dist}, \texttt{age}&	\texttt{dist}, \texttt{geo}&	0.3269\\
        genre&		subgenre&		\texttt{prev}, \texttt{dist}&		\texttt{prev}&		0.3287\\
        \hline
    \end{tabular}
    \label{table:twolevel_all}
\end{table}

We list the best performing depth-2 model for every combination of ($g_1,
g_2$) in Table~\ref{table:twolevel_all}, where we sort the performance by
misclassification error on development dataset. The feature \texttt{1}
indicates the feature of all ones (i.e.\ the intercept term).  Note that we decrease the
misclassification rate from 0.3212 to 0.3177 by adding an additional hierarchy
subgenre on top of author.  This improvement in predictive performance may
seem small, but in practice such improvements can translate to big impacts
when the corresponding models are deployed in commercial scale recommender
system applications~\citep{kramer2014experimental,kohavi2014seven}.  Thus, the
small improvement of the two-level model over the one-level model can be
meaningful.

The best two-level model is 
using subgenre and author as the two grouping factors; 
on subgenre level it uses \texttt{dist} and \texttt{age} as 
random effects features; on author level it uses \texttt{age} and 
\texttt{geo} as random features. 
It is a relatively simple model with competitive performance, 
and we will focus on this model throughout the rest of the paper.

\section{Results}
\label{sec:results}

\subsection{Performance}
\label{sec:comparison}

In Sec.~\ref{subsec:modelselection}, the model that gave the best prediction
performance on the development set used two levels of hierarchy, corresponding
to ``author'' and ``subgenere,'' with author nested within subgenre. For fixed
effect predictors, the model used an intercept along with \texttt{age},
\texttt{geo}, \texttt{prev}, and \texttt{dist}. For random effect predictors
at the first level in the hierarchy (subgenre), the model used an intercept
along with \texttt{dist} and \texttt{age}; at the second level in the
hierarchy (author) the model used an intercept along with \texttt{age} and
\texttt{geo}. Having selected the model, we will now evaluate its performance
on the held-out test set.

We fit the model to the training data set using our proposed moment-based
procedure \texttt{mhglm} along with two competing methods described in
Sec.~\ref{sec:simulation}, the \texttt{glmer} maximum likelihood procedure and
the \texttt{sgd} stochastic gradient descent $h$-likelihood-based procedure
, and we compare their prediction performances on the held-out test data set. 
We do not include the \texttt{glmer.split} method,
because \texttt{glmer} fails on the randomly splitted subsets due to sparsity.


\begin{table}[h!]
\centering
\caption{Misclassification Error and Running Time For Three Fitting Methods}
\begin{tabular}{ | c | r | c | r | } 
 \hline
 Fitting Method &  Error & Error 95\% Confidence Interval& Time (seconds)    \\ 
 \hline\hline
 \texttt{mhglm} &  $0.3262$    &   $\left[0.3189, 0.3335\right]$ &   55.14\\
\texttt{glmer} &  $0.3268$    &   $\left[0.3195, 0.3341\right]$   &   44790.23\\
\texttt{sgd}    & $0.3302$    &   $\left[0.3229, 0.3376\right]$ &   2022.35\\
\hline
\end{tabular}
\label{table:bookx_compare}
\end{table}

Table~\ref{table:bookx_compare} lists misclassification error, error's 95\%
confidence intervals, and running time for all three methods. All methods
have comparable prediction performance, with a misclassification rate of about
32.5\%. Our proposed procedure \texttt{mhglm} slightly outperforms the other
two methods in overall misclassification error, but the difference is not
statistically significant.  When we look at the running time, however,
\texttt{mhglm} is faster than \texttt{sgd} by a factor of 45, and faster than
\texttt{glmer} by a factor of 1000. Fitting the model using our proposed
method took under a minute; fitting using \texttt{sgd} took 33~minutes;
fitting using \texttt{glmer} took 12.4~hours.

The results in Table~\ref{table:bookx_compare} demonstrate two features of
the \texttt{mhglm} fitting procedure. First, the prediction performance is
comparable to that of the more established likelihood-based procedures.
Second, \texttt{mhglm} is faster than these methods by at least an order of
magnitude. This reduction in computation time enabled us to perform an
exhaustive model selection search over all 1- and 2-level models. For the
four predictors and the intercept, and for the 5 grouping levels, there were
$5 \cdot 2^4 = 80$ 1-level models and $10 \cdot 2^4 \cdot 2^4 = 2560$ 2-level
models. Extrapolating from the timing results in
Table~\ref{table:bookx_compare}, performing the search over these models
using \texttt{mhglm} took approximately 40 hours; using \texttt{sgd} or
\texttt{glmer}, the same search would take approximately 60~days or
3.75~years, respectively.

Our proposed fitting method has enabled us to perform an exhaustive search
over all 1- and 2-level hierarchical logistic regression models without
sacrificing prediction performance.



\subsection{Fitted model}
\label{sec:interpretation}

To gain some insight into the predictions made by our fitted model, we
investigate the empirical Bayes random effect estimates. Specifically, we
investigate the age random effects at the subgenre and author levels.

\begin{figure}
  \centering
  \begin{minipage}[b]{0.48\textwidth}
      \includegraphics[width=\textwidth]{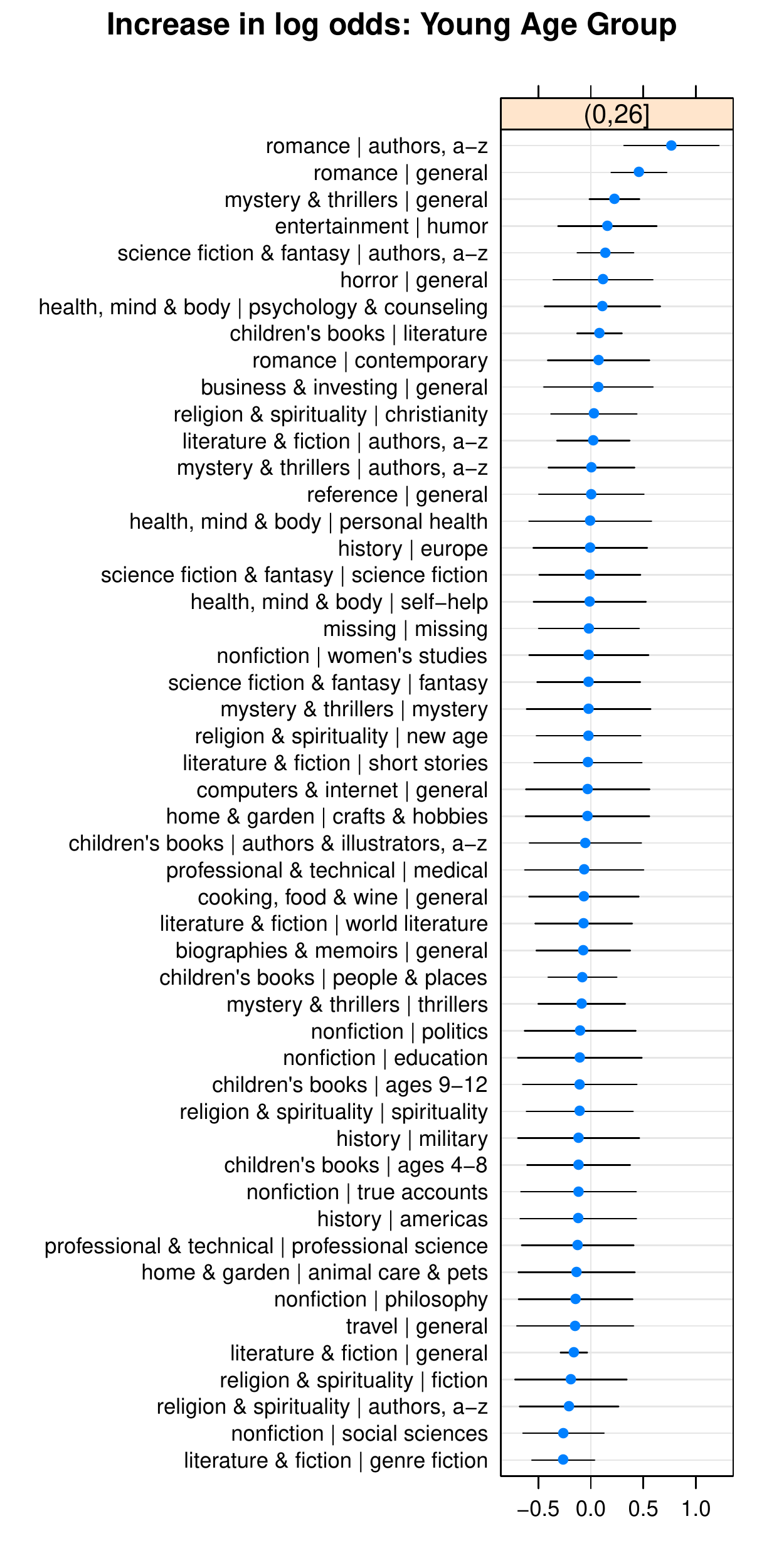}
  \end{minipage}
  \hfill
  \begin{minipage}[b]{0.48\textwidth}
      \includegraphics[width=\textwidth]{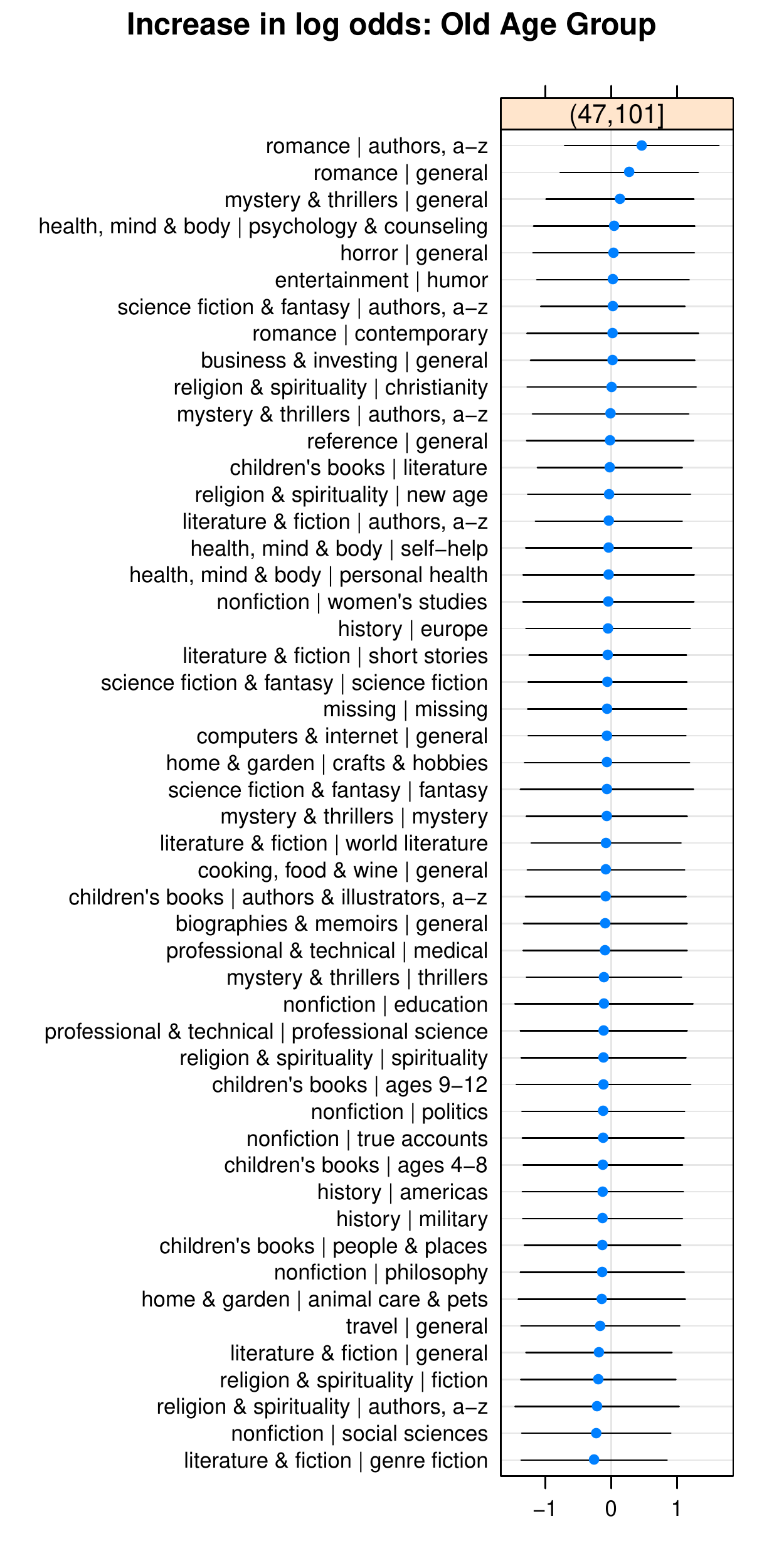}
  \end{minipage}
  \caption{Given book subgenre. Everything else remain the same, the increase in log odds
  if user is young (left panel) or old (right panel). Error-bars show the 
  $\pm$ estimated posterior standard deviation. Both figures show top 50 subgenres with most 
  ratings.}
\label{fig:caterpillar_cat2}
\end{figure}

In the context of the fitted model, given a book's subgenre we can compute the
increase in log odds of a user liking the book if we change the user's age
from ``missing'' to known while keeping all other predictors constant.  In
Fig.~\ref{fig:caterpillar_cat2} we show the change in log odds ($\pm1$
estimated posterior standard deviation) for young and old age groups, for the subgenres that have
most ratings.  For the old age group (47--101 years), the estimates have large
estimated posterior standard deviations across the subgenres listed, making it difficult to
identify a clear patter. For the young age group (0--26 years) there is some
weak but meaningful signal.  In this age group, there is a clear pattern in
which of the common subgenres the users like and don't like: ``romance'' is
their favorite subgenre, which has significantly higher random effects than
``literature \& fiction $|$ genre fiction'', their least favorite subgenre.

\begin{figure}
  \centering
  \begin{minipage}[b]{0.48\textwidth}
      \includegraphics[width=\textwidth]{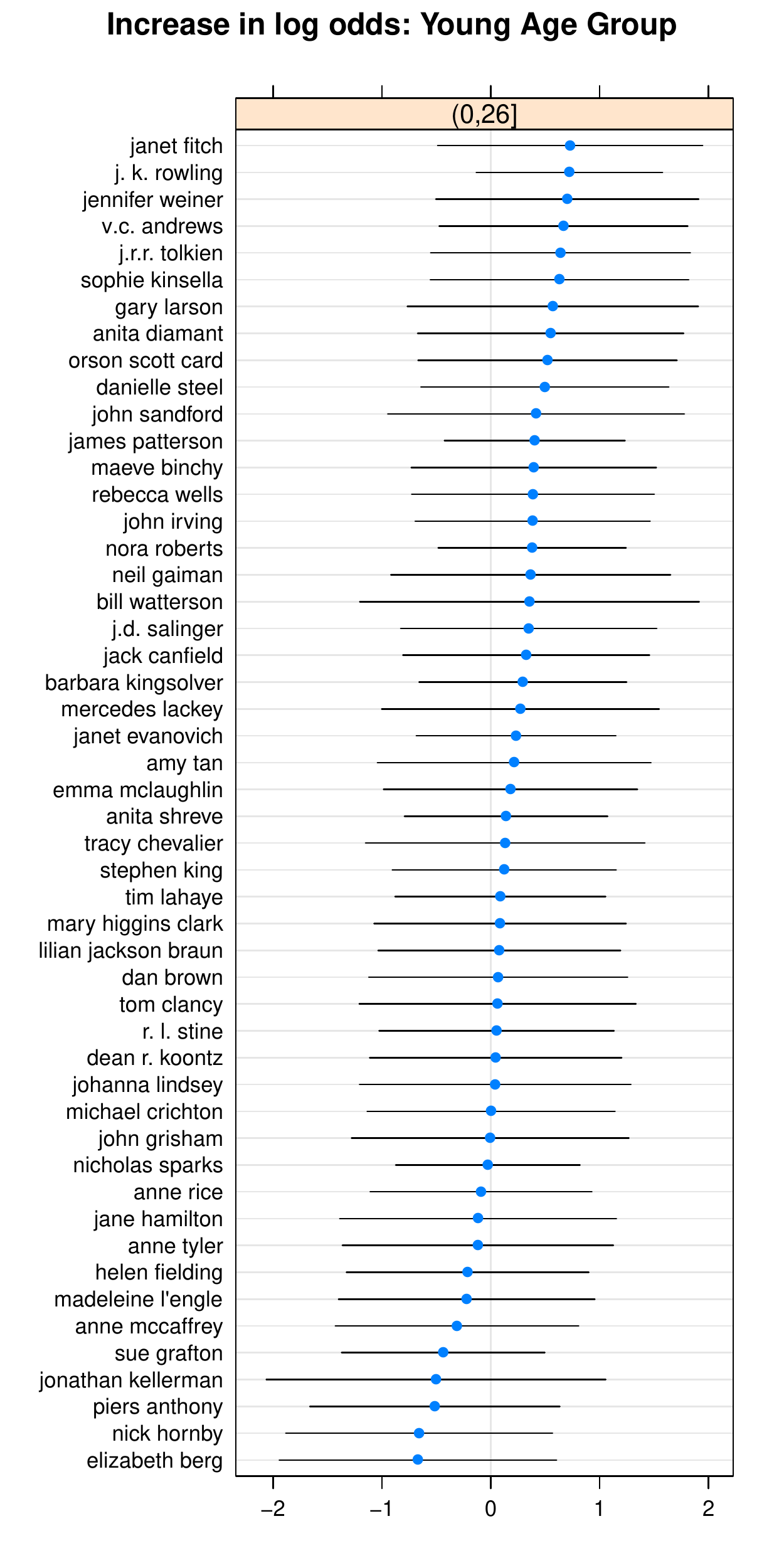}
  \end{minipage}
  \hfill
  \begin{minipage}[b]{0.48\textwidth}
      \includegraphics[width=\textwidth]{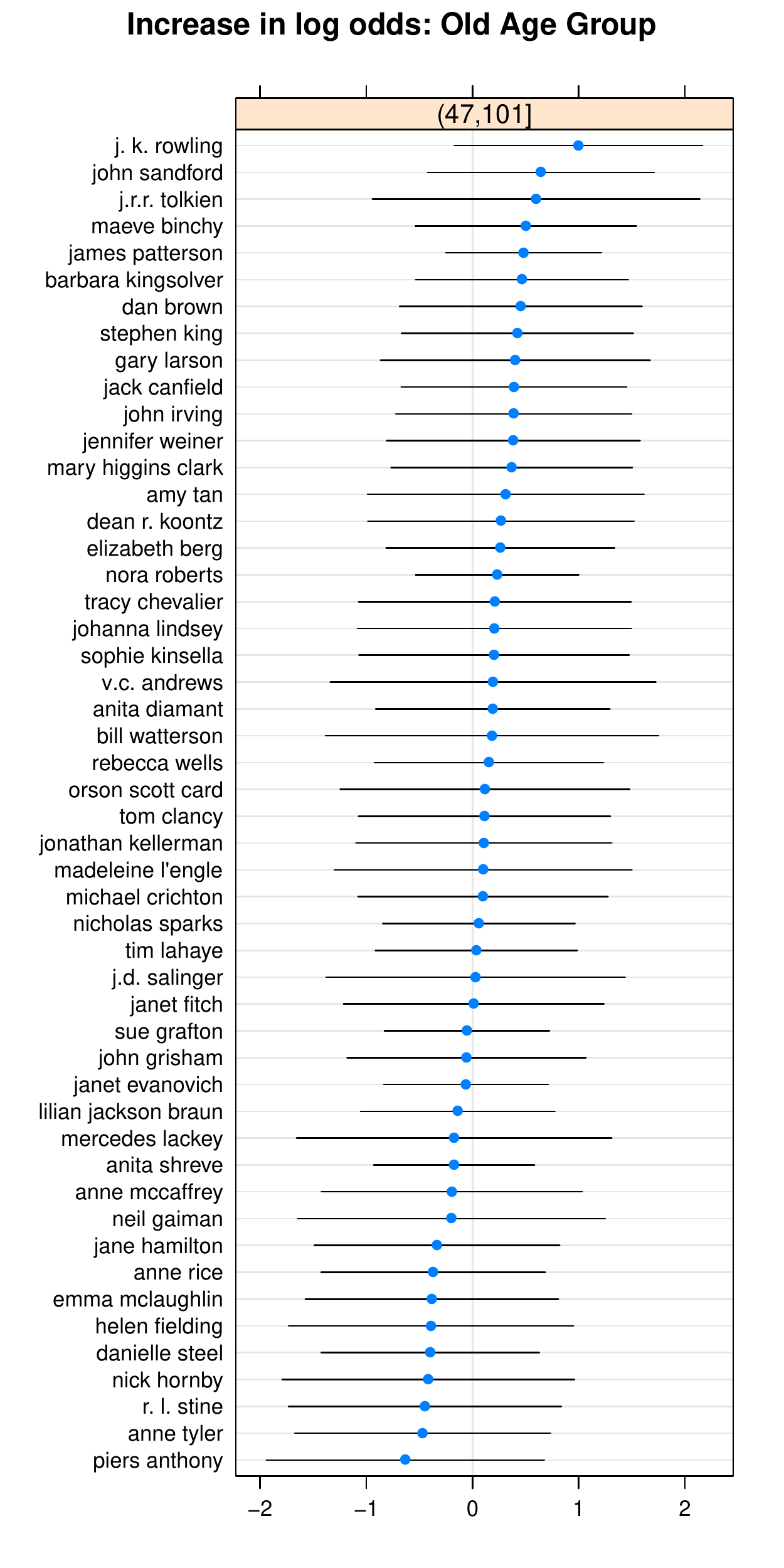}
  \end{minipage}
  \caption{Given book author. Everything else remain the same, the increase in log odds
  if user is young (left panel) or old (right panel). Error-bars show the 
  $\pm$ estimated posterior standard deviation. Both figures show top 50 authors with most 
  ratings.}
\label{fig:caterpillar}
\end{figure}

Next we perform a similar analysis, but on the second level of the hierarchy,
``author.''
For every author we compute the increase in
log odds of liking the book if we change the users' age from ``missing'' to
known while everything else remain the same. 
In Figure~\ref{fig:caterpillar} we show the increase in log odds 
($\pm$ estimated posterior standard deviation) for young and old age groups, 
for the authors that have most ratings.

We observe a few interesting patterns:
\begin{itemize}
    \item The estimated posterior standard deviations are much larger for random effects 
        on the second level (author), for both young and old age groups.
    \item Some authors are consistent across different age groups. 
        For instance, one would want to recommend J.~K.~Rowling 
        to both young and age groups. Meanwhile Nick Hornby and Piers Anthony
        are liked by neither groups.
    \item Some authors have quite different behaviors across age groups.
        For instance, Danielle Steel has positive log odds increase 
        if we know the user is within young age group, 
        but negative log odds increase if user is among old age groups.
        An opposite example is Elizabeth Berg: we will suffer a decrease
        in log odds if user is young, meanwhile log odds will increase 
        if the user is old.
\end{itemize}

\noindent
The size of the estimated posterior standard deviations make clear that these associations are
weak. Still, as demonstrated in
Secs.~\ref{subsec:modelselection}~and~\ref{sec:comparison}, there is enough signal
in them to translate to a meaningful reduction in misclassification rate on
the held-out development and test sets.

\section{Discussion}
\label{sec:discussion}

The appeal of the deeply-nested hierarchical model is that it facilitates
information sharing across subtrees at all levels of the hierarchy. Nodes with
abundant data effectively have their random effects estimated using only data
at the leaves descending from them. Nodes with little or moderate data, however,
benefit by having their estimated coefficients (random effects) shrunk towards
the global mean. In our book recommendation application, we have demonstrated
this advantage by showing that using two levels of hierarchy (author and
subgenre) delivers increased prediction performance than using one or no
levels.

The main hurdle in deploying hierarchical models in recommender systems
applications like ours and other contexts of similar scale is that the time
required to fit these models can be prohibitive. \citet{Perr16} extended
a method original due to \citet{Coch37} and proposed a
partial solution to this problem, but his procedure is limited to single-level
hierarchical models. Here, with our proposed \texttt{mglhm} method we have
shown how to fit a hierarchical model of arbitrary depth by repeatedly
applying the single-level fitting procedure to prune the leaves of the
hierarchy. We then showed how to propagate the estimates at the root of the
hierarchy down through the nodes in the hierarchy to refine the random effect
estimates.

In our book recommendation application, our proposed fitting procedure was
faster than stochastic gradient descent by a factor of 45, and faster than the
likelihood-based \texttt{glmer} procedure by a factor of 1000. This increase
in computational speed enabled us to perform an exhaustive model selection
search over all one- and two-level models, reducing the overall computation
time from about 60 days using \texttt{sgd} (or 3.75 years using
\texttt{glmer}) to about 40 hours. As our simulations in
Sec.~\ref{sec:simulation} demonstrated, the tradeoff in deploying our method
is reduced statistical efficiency and prediction performance. However, in our
application, the loss in prediction performance was negligible.

Although our motivation was a book recommendation system, our proposed fitting
procedure is general enough to handle hierarchical generalized linear models
of arbitrary depth. We have incorporated our implementation of this procedure
into the \texttt{mbest} R package, available on the Comprehensive R Archive
Network (CRAN). The interface in this implementation is flexible enough to
handle any deeply nested hierarchical generalized linear model.

\appendix

\section{Data description}
\label{app:bookxdata}
\subsection{Overview}

The Book-Crossing dataset introduced in Section~\ref{sec:intro} contains
433,671 numerical ratings of 185,973 books from 77,805 users~\citep{Zieg05}.  
Each rating consists of a book id (ISBN), a user id, and a numerical score between
1 and 10, where 1 indicates extreme negative and 10 indicates extreme positive
sentiment. We binarize the ratings so that ratings equal or above $8$ are
considered positive and ratings below $8$ are considered negative.
The threshold $8$ is chosen such that the two classes have comparable
number of samples.
We have user demographic information including age and location;
Section~\ref{bookx:userdemo} reports some descriptive statistics about these
features. We also know the book authors and titles.

We augment the book meta-data with a genre hierarchy scraped from
Amazon.com by~\citet{Mcau15}. In this meta-data, book titles are nested within authors
within sub-subgenres within subgenres within genres. If the same author writes
titles in multiple sub-subgenres, we treat the author as multiple, separate entities.
Section~\ref{bookx:hier} describes the hierarchy in meta-data.

In the raw dataset, more than half of the ratings cannot be matched
to Amazon meta-data. Dealing with this missing data is beyond the scope of the
present treatment, so we remove samples with missing ratings or unmatched book
ids from consideration. This leaves us with 157,638 ratings of 38,659 books
from 38,085 users.

\subsection{User demographic features}
\label{bookx:userdemo}

The reported user age is a continuous variable, ranging from 15 to 100.  
The mean and standard deviation of user age are 36.4 and
12.6 respectively.  Table~\ref{table:age} shows the number of users and
ratings from each age range. 

\begin{table}[h!]
\centering
\caption{Number of Users \& Ratings from Each Age Range.}
\begin{tabular}{ | c | r | r | } 
 \hline
 Age Interval & \# Users & \#  Ratings    \\ 
 \hline\hline
$\le$ 20&2,664&8,752\\
(20,30]&6,011&30,820\\
(30,40]&5,906&32,252\\
(40,50]&3,765&20,539\\
(50,60]&2,609&11,961\\
(60,70]&952&2,997\\
$>$ 70 & 297& 992\\
\hline
\end{tabular}
\label{table:age}
\end{table}

Most of the ratings comes from young or middle-aged users,
which makes it easier to estimate and predict for users from those age ranges.

User's location information is reported as his city, state, country, and
continent.  Table~\ref{table:geocountry} reports the number of users and
ratings from the 10 most-represented countries, and
Table~\ref{table:geocontinent} reports the same information for each
continent.

\begin{table}[h!]
\parbox{.45\linewidth}{
\centering
\caption{Top 10 Countries With Most Ratings.}
\begin{tabular}{ | c | r | r | } 
 \hline
 Country &\# Users & \# Ratings  \\ 
 \hline\hline
USA&29,042&120,201\\
Canada&3,619&14,592\\
United Kingdom&989&3,622\\
Australia&632&2,067\\
Portugal&181&1,490\\
Germany&381&1,189\\
Spain&187&1,008\\
Malaysia&111&964\\
Netherlands&178&638\\
New Zealand&148&563\\
\hline
\end{tabular}
\label{table:geocountry}
}
\hfill
\parbox{.45\linewidth}{
\centering
\caption{Number of Users \& Ratings from Each Continent.}
\begin{tabular}{ | c | r | r | } 
 \hline
 Continent & \# Users & \# Ratings  \\ 
 \hline\hline
 North America&32,722&135,059\\
Europe&2,632&10,122\\
Oceania&780&2,630\\
Asia&430&2,272\\
South America&68&210\\
Africa&36&87\\
\hline
\end{tabular}
\label{table:geocontinent}
}
\end{table}

We can see that the vast majority of the ratings ($85\%$) are from North
America, with Europe, the next-most-represented country, receiving only
$6\%$ of ratings. This indicates that it's quite difficult to accurately 
estimate and predict for users from other than these two continents.

\subsection{Book hierarchy}
\label{bookx:hier}

Every book is nested under a deep hierarchy:
\begin{center}
  genre $\triangleright$ subgenre $\triangleright$ sub-subgenre
  $\triangleright$ author $\triangleright$ title.
\end{center}
For example, the book \emph{Harry Potter and the Chamber of Secrets} is nested
as \emph{Children's Books} $\triangleright$ \emph{Literature} $\triangleright$
\emph{Science Fiction, Fantasy, Mystery \& Horror} $\triangleright$
\emph{J.~K.~Rowling} $\triangleright$ \emph{Harry Potter and the Chamber of
Secrets}. Figure~\ref{fig:2lpie} displays all hierarchies on the first two levels.


For modeling purposes, we only chose two out of five hierarchies. 
We omit the intermediate levels and use simplified
hierarchy of subgenre $\triangleright$ author.  
Our first level of hierarchy subgenre has 1,344 groups,
which captures the necessary amount of diversity across books
using a reasonable amount of groups.
We use author (nested under subgenre) as the second level 
of hierarchy, which has 27,360 groups. We use book author instead of 
book title as the second level hierarchy, since the Book Crossing dataset 
is very sparse, such that most books has only a few number of ratings. 
Hierarchical models will not work well if most groups have very few samples,
which shrinks the overall results towards that of a simple ``global'' model. 

\begin{figure}[h!]
  \centering
  \begin{minipage}[b]{0.48\textwidth}
    \includegraphics[width=\textwidth]{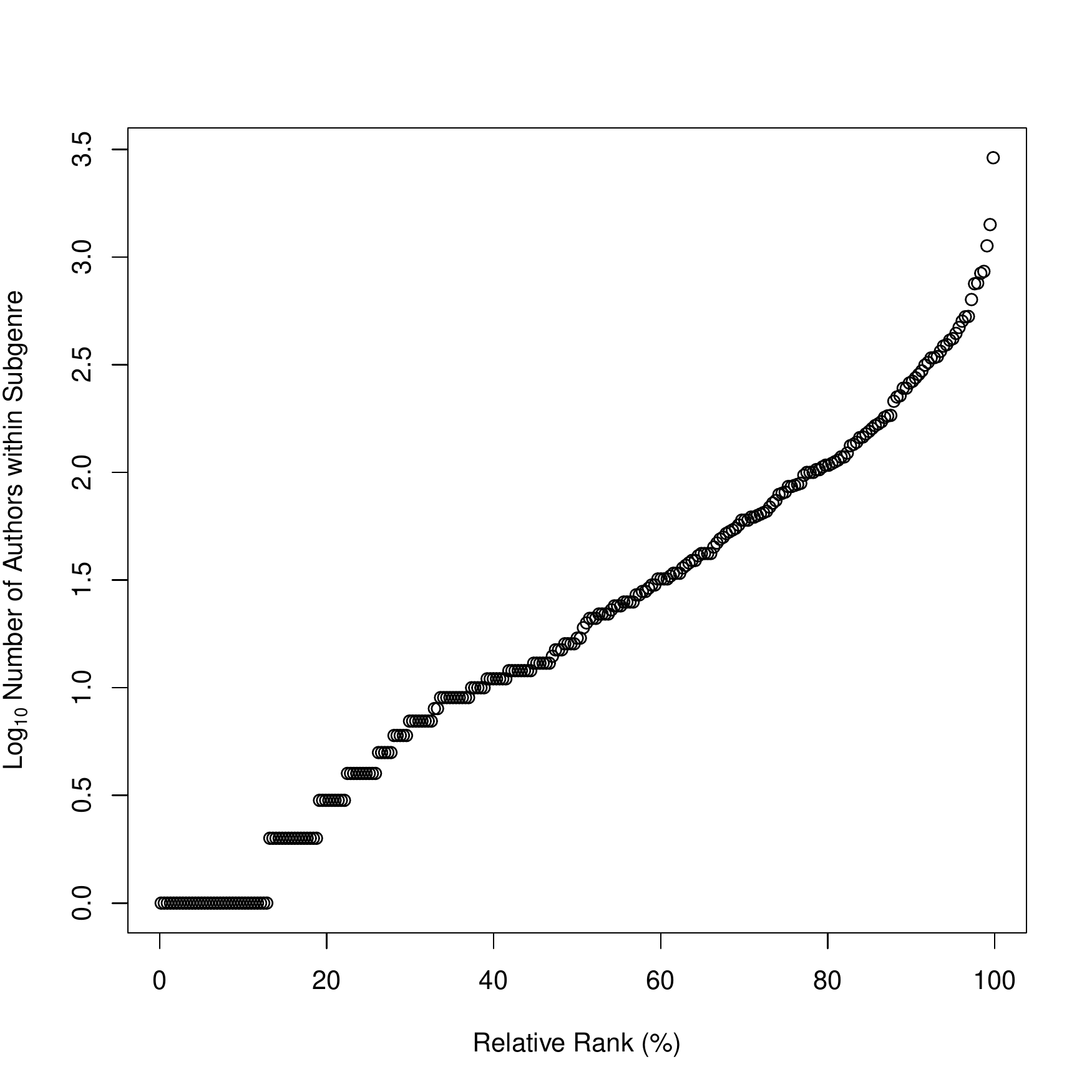}
  \end{minipage}
  \hfill
  \begin{minipage}[b]{0.48\textwidth}
    \includegraphics[width=\textwidth]{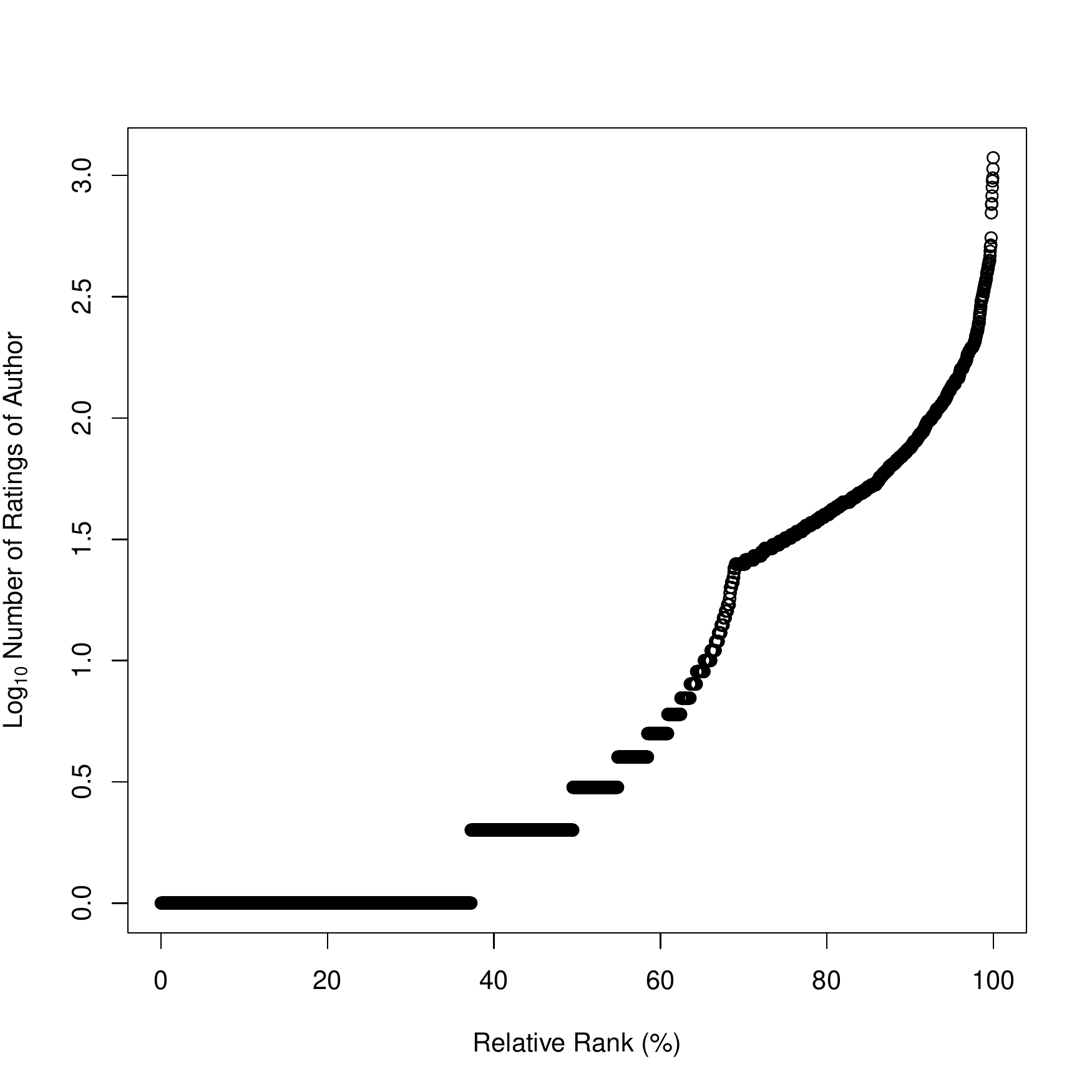}
  \end{minipage}
  \caption{Left panel: Quantile Plot of $Log_{10}$ Number of Authors within Subgenres. 
  Right panel: Quantile Plot of $Log_{10}$ Number of Ratings of Authors.}  
\label{fig:cum}
\end{figure}

Even for these carefully chosen hierarchies, the distribution of subgroups
and samples are still highly skewed.
We can see this skewness in Figure~\ref{fig:cum}, which plots the quantiles
of the number of authors per subgenre (left panel) and
the number of ratings per author (right panel). 
Both plots are on $log_{10}$ scale.
50\% of subgenres have fewer than 17 authors, and 90\% of subgenres
have fewer than 261 authors.  At the other extreme, the largest subgenre
(Literature \& Fiction $\triangleright$ General)
has 2893 authors.  The distribution of ratings among authors are highly skewed
as well: 50\% of authors have only 1 rating, 90\% of authors have less than 9
ratings, meanwhile the mostly rated author (Sue Grafton) received 1183
ratings.

Hierarchical models gain its predictive power by pooling information across
groups. The existence of large numbers of small groups will make learning
model parameters as and making good predictions difficult.

\section{Two-level linear model simulations}
\label{app:twolevel-linear}

Here we perform a simulation study similar to the two-level logistic
regression model study described in Sec.~\ref{sec:simulation}, but using
a two-level linear regression model instead.

\begin{figure}
 	\centering
 	\includegraphics{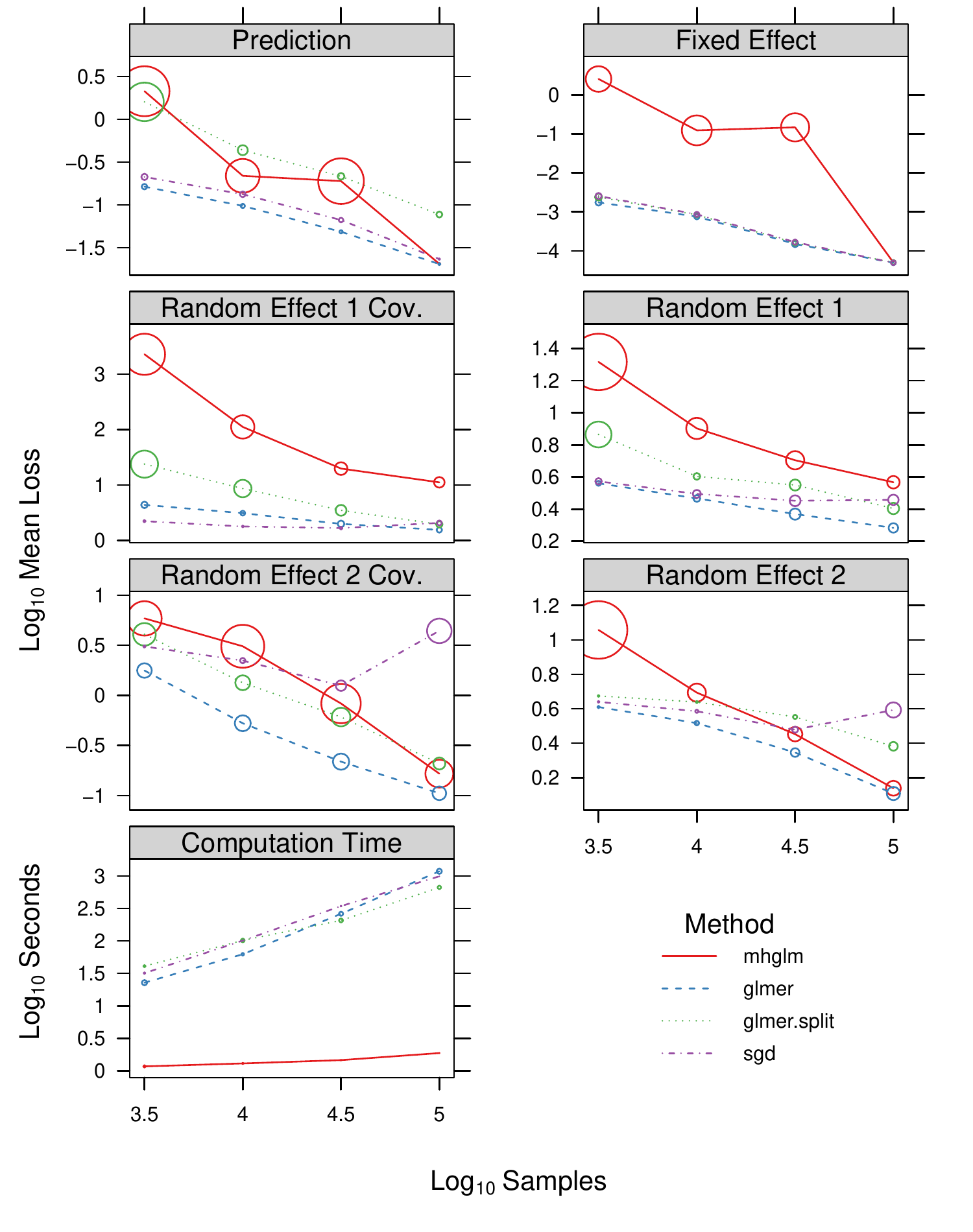}
 	\caption{Performance for the multilevel linear regression model.
	Circle radii indicate one standard error along $y$-axis (absent when smaller than line width)}
 	\label{fig:sim-gaussian}
\end{figure}

With all other simulation parameters drawn as described in
Sec.~\ref{sec:simulation}, in the linear regression setup we draw response~$k$
from a normal distribution with mean $\mu_k=x_k^T\beta+z_k^T(u_{i}+u_{ij})$ 
and variance $\phi=1$ whenever sample~$k$ belongs to leaf $ij$.
We again compare our procedure with those three methods. We use the same loss for fixed and random effects, 
as well as the random effect covariance. For prediction loss, we use the mean squared error:
\begin{equation*}
N^{-1}\sum_{k=1}^{N}\phi^{-1} (\mu_k - \h\mu_k)^2
\end{equation*}
where $\mu_k=x_k^\T\beta+z_k^\T(u_{i}+u_{ij})$ and 
$\h\mu_k=x_k^\T\h\beta+z_k^\T(\h u_{i}+\h u_{ij})$.

Figure~\ref{fig:sim-gaussian} shows the mean loss, averages over $20$ replicates, with
circle radii indicating standard errors along the vertical axes. For moderate to large 
sample sizes, there is a noticeable but decreasing difference between the proposed method and 
other maximum likelihood based estimators. However the proposed method still appears to be consistent. 
In terms of computation time, this method again has improvement by factor ranging from 
 $100$ to $1000$, and the factor appears to grow exponentially as sample sizes increase.


\bibliographystyle{imsart-nameyear}
\bibliography{mlevels}

\end{document}